\documentclass[aps,prl,twocolumn,floatfix]{revtex4}
\pdfoutput=1
\usepackage{graphicx}
\usepackage{hyperref}

\usepackage{amssymb,amsfonts,amsmath}

\newcommand{\ket}[1]{| #1 \rangle}

\begin{document} 

\title {Chaotic Einstein-Podolsky-Rosen pairs,
measurements and time reversal}
\author{Klaus M. Frahm and Dima L. Shepelyansky}

\affiliation{\mbox{Laboratoire de Physique Th\'eorique,
Universit\'e de Toulouse, CNRS, UPS, 31062 Toulouse, France}}

\date{June 6, 2021}


\begin{abstract}
{\bf Abstract:} 
We consider a situation when evolution
of an entangled Einstein-Podolsky-Rosen (EPR) pair
takes place in a regime of quantum chaos
being chaotic in the classical limit.
This situation is studied on an example 
of chaotic pair dynamics described by 
the quantum Chirikov standard map.
The time evolution is reversible 
even if a presence of small errors
breaks time reversal of classical dynamics due to
exponential growth of errors induced by exponential chaos
instability. However, the quantum evolution
remains reversible since a
quantum dynamics instability exists only on 
a logarithmically short Ehrenfest time scale.
We show  that due to EPR pair entanglement 
a measurement of one particle at the moment of time reversal
breaks exact time reversal of another particle 
which demonstrates only an approximate time reversibility.
This result is interpreted in the framework of the Schmidt decomposition 
and Feynman path integral formulation of quantum mechanics.
The time reversal in this system has already been 
realized with cold atoms in kicked optical lattices
in absence of entanglement and measurements.
On the basis of the obtained results
we argue that the experimental investigations of
time reversal of chaotic EPR pairs is within
reach of present cold atom capabilities.
\end{abstract}

\maketitle

\section{I. Introduction} 
\label{sec1}
The fundamental work of
Einstein-Podolsky-Rosen (EPR) \cite{epr} on a distant entanglement of 
a pair of non-interacting distinguished 
particles and its effects on measurements
is now at the foundations of long-distance quantum communications.
The entanglement concept coined by Schr\"odinger \cite{schrodinger}
with a gedanken experiment of a cat, dead or alive,
becomes a resource of modern quantum computations \cite{chuang,karol}.
An impressive modern progress of quantum information, 
computation and communication
is described in \cite{deutsch}. 

An overview of various experimental
realizations of EPR pairs is given in \cite{eprexprev}.
Various forms of propagating EPR pairs have been studied
experimentally but in its main aspect the propagation
of EPR pairs was rather simple being similar
to propagation on a line and always being integrable.
Here we consider theoretically a  situation when 
two non-interacting but entangled  particles of an EPR pair propagate
in a regime of quantum chaos \cite{haake}.
In the classical limit a dynamics of these particles is chaotic 
being characterized by an exponential local divergence
of trajectories with a positive Kolmogorov-Sinai entropy $h$
\cite{arnold,sinai,chirikov1979,lichtenberg}. The exponential instability
of chaotic dynamics leads to exponential growth of round-off errors
and breaking of time reversibility of classical evolution
described by reversible equations of motion.
Thus chaos resolves the famous Loschmidt-Boltzmann dispute
on time reversibility and emergence of statistical
laws from reversible dynamical equations \cite{boltzmann1,loschmidt,boltzmann2}
(see also \cite{mayer}). 
Prior to classical chaos theory the problem of time reversal of 
laws of nature was also discussed by such leading scientists as 
Schr\"odinger \cite{schrodinger1931} 
(see English translation and overview in \cite{schrodtrans}) and 
Kolmogorov \cite{kolmogorov}.

However, in quantum mechanics a chaotic mixing in a phase-space cannot go 
down to exponentially small scales being restricted by a quantum scale of the 
Planck constant $\hbar$. Thus in the regime of quantum chaos
an exponential instability exists only during a logarithmically 
short Ehrenfest time
scale $\tau_E \sim |\ln \hbar|/h$ \cite{chi1981,dls1981,chi1988,ehrenfestime}
(here $\hbar$ is a dimensionless effective Planck constant 
related to typical quantum numbers). 
Due to the absence of exponential instability on
times beyond $\tau_E$ the quantum evolution remains reversible 
in presence of quantum errors in a drastic difference from
the classical dynamics as it was demonstrated
in \cite{dls1983} for the quantum Chirikov standard map,
also known as a kicked rotator \cite{chi1981,chi1988,stmap}.
This system has been experimentally realized with cold atoms
in kicked optical lattices and in particular 
the quantum dynamical localization of
chaotic diffusion has been observed in these experiments
\cite{raizen,garreau}.  This dynamical localization of chaotic diffusion
appears due to quantum interference and is analogous 
to the Anderson localization \cite{anderson}
of electron diffusion in disordered solids (see e.g. \cite{fishman1,fishman2,dls1987}).

In \cite{martin} it was shown that the time evolution of cold atoms in 
kicked optical lattices, described by the quantum Chirikov standard map,
can be reversed in time in the regime of quantum chaos.
This proposal was indeed experimentally realized by the Hoogerland 
group \cite{hoogerland}.
Thus this system represents an efficient experimental platform which
allows to investigate nontrivial effects of quantum mechanics, localization,
chaos and time reversal.

In this work we investigate the properties of chaotic EPR pairs
evolving in this fundamental system of quantum chaos
and show that a measurement of one of the entangled particles
breaks exact time reversal of the other particle but 
preserves its approximate time reversibility.
We explain this unusual effect on the basis of 
the Schmidt decomposition \cite{schmidt}
(see also the review \cite{fedorov} and Refs. therein)
and the  Feynman path integral formulation of quantum mechanics \cite{feynman}.

This article is composed as follows: the model is described in Section II,
the results are presented in Section III
and the discussion and conclusion are given in Section IV;
additional Figures and data are given in Appendix.

\section{II. Model description} 
\label{sec2}

The classical dynamics of one particle is described by
the Chirikov standard map \cite{chirikov1979}:
\begin{equation}
\label{stmap}
\bar{p} = p + k \sin{ x} \; , \;\; 
\bar{x} = x + T \bar{p} \; .
\end{equation}
Here $x$ represents the position of an atom
in an infinite $x-$axis of the kicked optical lattice,
or  a cyclic variable $0 \leq x < 2\pi$
for the case of the kicked rotator; $p$ is the momentum of a particle.
The bars denote the new values of variables after one iteration of this 
symplectic map. 
The physical process described by this map corresponds to
a sharp change of momentum, generated 
e.g. by a kick  of the optical lattice \cite{raizen,garreau},
followed by a free particle propagation during a period $T$ between kicks.
The classical dynamics depends on a single chaos parameter
$K=kT$ with a transition from integrability to unlimited chaotic
diffusion in momentum for $K > K_c =0.9715...$ \cite{chirikov1979,lichtenberg}.
The system dynamics is reversible in time, e.g. by
inverting all velocities in a middle of free rotation between two kicks.

Inside a chaotic component the dynamics is characterized by 
an exponential divergence of trajectories with the positive
Kolmogorov-Sinai entropy $h$. For $K>4$ the measure of stability
islands is small and we have $h  \approx \ln(K/2)$ \cite{chirikov1979}.
For $K > K_c$ the dispersion of momentum
grows diffusively with time $\langle(\Delta p)^2\rangle = D t$ 
with a diffusion coefficient
$ D \approx k^2/2$ (see more details in  \cite{chirikov1979,dls1987}).
Here and below the time $t$ is measured in number of map iterations.
The map captures a variety of universal features
of dynamical chaos and appears in the description
of various physical systems \cite{stmap}. 

The quantum evolution of the state $\ket{\psi}$ over a period is given 
by a unitary operator
 $\hat{U}$ \cite{chi1981,chi1988}:
\begin{eqnarray} 
\label{qmap}
\ket{\bar{\psi}} = \hat{U} \ket{\psi} = 
e^{-iT\hat{p}^2/2} e^{-ik\cos{\hat{x}}} \ket{\psi} \; .
\end{eqnarray} 
Here the momentum $p$ is measured in recoil units of optical lattice with 
$\hat{p}=-i \partial / \partial x $. Thus $T=\hbar$ plays the role of
an effective dimensionless Planck constant and the classical limit
corresponds to $T=\hbar \rightarrow 0$, $k \rightarrow \infty$,
$K=kT = const$. Due to the periodicity of the optical lattice potential
the momentum operator $\hat{p}=-i \partial / \partial x $ has eigenvalues
$p=n+\beta$ where $n$ is an integer and
$\beta$ is a quasimomentum conserved by the kick potential 
($0 \leq \beta < 1$). The value $\beta=0$
corresponds to the case of a kicked rotator
with a wave function (in position representation) 
$\psi(x)=\langle x\ket{\psi}$ being 
periodic on a circle $\psi(x+2\pi)=\psi(x)$.
In this case the free rotation correspond (in momentum representation) 
to the phase shift
${\bar {\psi}}_{n,0} = \exp(-iTn^2/2) \psi_{n,0}$ with 
$\psi_{n,\beta}=\langle p\ket{\psi}$ being the wave function 
(in momentum representation) at $p=n+\beta$. 
Irrational values of $\beta$ appear for
a particle propagation on an infinite $x$-axis;
here $\beta$ is conserved
and a free propagation of the momentum wave function $\psi_{n,\beta}$
gives the phase shift ${\bar {\psi}}_{n,\beta} = 
\exp(-iT(n+\beta)^2/2)\,\psi_{n,\beta}$.
The effects of quantum interference lead to dynamical localization
of chaotic diffusion on a time scale $t_D \approx D/\hbar^2 \gg \tau_E$
and an exponential localization of quasienergy eigenstates
with a localization length $\ell = D/(2 \hbar^2) \approx k^2/4$ \cite{dls1987,chi1988}.

In \cite{martin} it was pointed that the time reversal of a quantum evolution
after $t_r$ map iterations
can be realized by using a period between kicks
being $T=4\pi+\epsilon$ for $t \leq t_r$
and $T'=4\pi - \epsilon$ for $t_r < t \leq 2 t_r$.
Also the time reversal is done at the middle of the free propagation 
after $t_r$ kicks (it is convenient to use a symmetrized scheme with a
half-period of free  rotation then kick and then again 
a half-period of free propagation).
The inversion of kick amplitude
$k \cos x \rightarrow - k \cos x$ can be realized 
by a $\pi$-translational shift of the optical lattice potential.
Such a time reversal is exact for $\beta=0$ (kicked rotator case)
and it also works approximately for small $\beta$ values
in the case of the kicked particle \cite{martin}. 
The time reversal for cold atoms in a kicked optical lattice
was experimentally demonstrated in \cite{hoogerland}.

Here we consider the time reversal of two
non-interacting distinguished particles being in an initial
entangled state. We concentrate our analysis on the case when
both particles evolve in the regime of quantum chaos.
Thus we have the new case of chaotic EPR pairs. Following (\ref{qmap})
the evolution of the two particle state $\ket{\psi}$ 
(with wave function $\psi(x_1,x_2)=\langle x_1,x_2\ket{\psi}$) 
of such pairs is given by the quantum map
\begin{eqnarray} 
\label{qmappair}
\ket{\bar{\psi}} = (\hat{U}_1\otimes \hat{U}_2) \ket{\psi}   \; ,
\end{eqnarray} 
where $\hat{U}_1$ and  $ \hat{U}_2$ are one time period
evolution operators for the first and second particle. 
In absence of interactions between particles
the entropy of entanglement $S$ is preserved during this time
evolution. It is convenient to use the Schmidt decomposition 
\cite{schmidt,fedorov} 
for an initial entangled state
\begin{eqnarray} 
\label{schmidt}
\ket{\psi}=\sum_{i=1}^m \alpha_i \ket{u_i}\otimes\ket{v_i}
\end{eqnarray} 
where $\ket{u_i}$, $\ket{v_i}$ are one-particle states satisfying 
the orthogonality relations: 
$\langle u_i\ket{u_j}=\langle v_i\ket{v_j}=\delta_{ij}$. 
The number $m$ of Schmidt components can be up to $m=N$ if $N$ is the 
dimension of the one-particle Hilbert space. However, for ``less'' entangled 
states $m$ may be smaller and in this work we will consider the case of 
$m=2$.
The entropy of entanglement is then 
given by (see e.g. \cite{chuang,fedorov}):
 \begin{eqnarray} 
\label{entropy}
S = -Tr(\rho_1 \log_2 \rho_1) = - \sum_i |\alpha_i|^2 \log_2  |\alpha_i|^2 \; ,
\end{eqnarray} 
where $\rho_1$ is a reduced density matrix of first particle
obtained by a trace taken over the second particle.
During the time evolution of EPR pair given by (\ref{qmappair})
the wave functions of each particle evolve independently
with $\ket{u_i(t)} = {\hat{U}_1^t} \ket{u_i(t=0)}$
and  $\ket{v_i(t)} = {\hat{U}_1^t} \ket{v_i(t=0)}$.
Thus the coefficients $\alpha_i$ of the Schmidt decomposition and the 
entropy of entanglement $S$ remain unchanged.

However, since the particles are entangled
a measurement of the second particle after the time $t_r$ affects the 
wave function of first particle and thus the time reversal evolution of this 
particle is modified so that the exact time reversibility
is broken by the measurement. Nevertheless, we will see that still there is
an approximate time reversal of the first particle.
We describe in detail this effect in the next section.

\section{III. Time evolution of chaotic EPR pairs} 
\label{sec3}

The numerical simulations of the quantum map (\ref{qmap}),~(\ref{qmappair})
are done in a usual way \cite{chi1981,chi1988} 
by using the fact that the free propagation and the kick 
are diagonal in the momentum and coordinate representations respectively.
Concerning the eigenphases $T n^2/2$ of the free propagation operator 
we mention an important technical detail: we compute these phases 
for $n=-N/2,\,\ldots,\,N/2-1$ 
(with $N$ being the dimension of the one-particle Hilbert space) 
and the values for $n<0$ are stored at the positions $N-n$ while the 
values for $n\ge 0$ are stored at positions $n$. In this way 
if the initial states are localized close to small values of $n\approx 0$ 
(or $n\approx N$ which is topologically close to $n\approx 0$ due 
to the periodic boundary conditions) and 
if during the time evolution the states do not touch the borders at 
$n\approx \pm N/2$ the results are independent of the exact choice $N$ 
provided $N$ is sufficiently large. In other words the momentum phases 
exhibit a smooth transition between $n\approx 0$ and $n\approx N$ according 
the quadratic formula while at the ``system border'' $n\approx N/2$ 
this transition is not smooth. Otherwise, if the phases were naively 
computed for $n=0,\,\ldots,N-1$ according to the quadratic formula the 
results would depend in a sensitive way on $N$ even if the states remain 
localized close to $n\approx 0$ since the eigenphases for $n\approx N$ 
would be very different.

The transitions from one representation (momentum or position) to another and
back are done with the Fast Fourier Transform (FFT).
Furthermore, we chose the quantum map to be directly symmetric in time and 
therefore we present it as a half period of free propagation 
(using the operator $\hat U_{\rm half,free}=e^{-iT\hat{p}^2/4}$)
followed by the kick (using $\hat U_{\rm kick}=e^{-ik\cos{\hat{x}}}$) 
and then again a half period of free propagation (using 
$\hat U_{\rm half,free}$).
Furthermore, in order to have an exact mathematical equivalence 
between the two cases $T=4\pi+\epsilon$ and $T=\epsilon$ 
(at $\beta=0$) we also apply for 
the first case to the initial states (given below for the different cases
we consider) 
an initial half period of free propagation 
with $T=4\pi$ (which provides an additional phase factor $(-1)^{n_1+n_2}$ 
in momentum representation). We have numerically verified that this 
equivalence is indeed valid. 

We consider in detail 3 specific cases:
A) kicked rotator case with a moderate dimensionless effective Planck constant
$\hbar_{\rm eff}=\epsilon = T -4\pi < 1$ and a wavefunction periodic 
on the $2\pi$-circle (i.e. integer values of $p_i=n_i$ with $\beta_i=0$ 
and $i=1,2$ for both particles);  
B) same case but taken in the deep semiclassical regime with 
$\hbar_{\rm eff}  \ll 1$; C) the case of kicked particles propagation
on an infinite (or quasi-infinite) line at moderate $\hbar_{\rm eff}$
that corresponds to the case of cold atoms in a kicked
optical lattice \cite{raizen,garreau,hoogerland} composed of $L$ periods 
such that $x\in[0,\,2\pi L$. 
The total computational basis size for one particle,
used in the numerical simulations,
was changing from $N=1024$ up to $N=2^{22}$,
depending on the choice of A), B), C) and insuring 
that the basis size does not affect the obtained results.
For two particles the size of the Hilbert space is
$N_H=N^2$. For moderate values of $N$ (e.g. up to $N=2^{12}$ 
in cases A and B)
we used the whole basis with $N_H$ states
using two-dimensional (2D) FFT
transitions between momentum and coordinate 
representations in (\ref{qmappair}).
For larger $N$ values we used the fact that 
the Schmidt decomposition (\ref{schmidt}) has coefficients $\alpha_i$
being unchanged during the time evolution
so that we propagate independently each particle
and use the Schmidt entangled EPR wavefunction
for a measurement of the second particle
at the time moment $t_r$ 
and backward propagating only the first particle
after measurement. 
We checked, for $N \leq 2^{12}$, that these two numerical 
methods of time evolution simulation
give the same results up to the computer numerical accuracy.
Some additional details about numerical simulations 
and Figures are given in the Appendix.

\begin{figure}[h!]
\begin{center}
\includegraphics[width=0.45\textwidth]{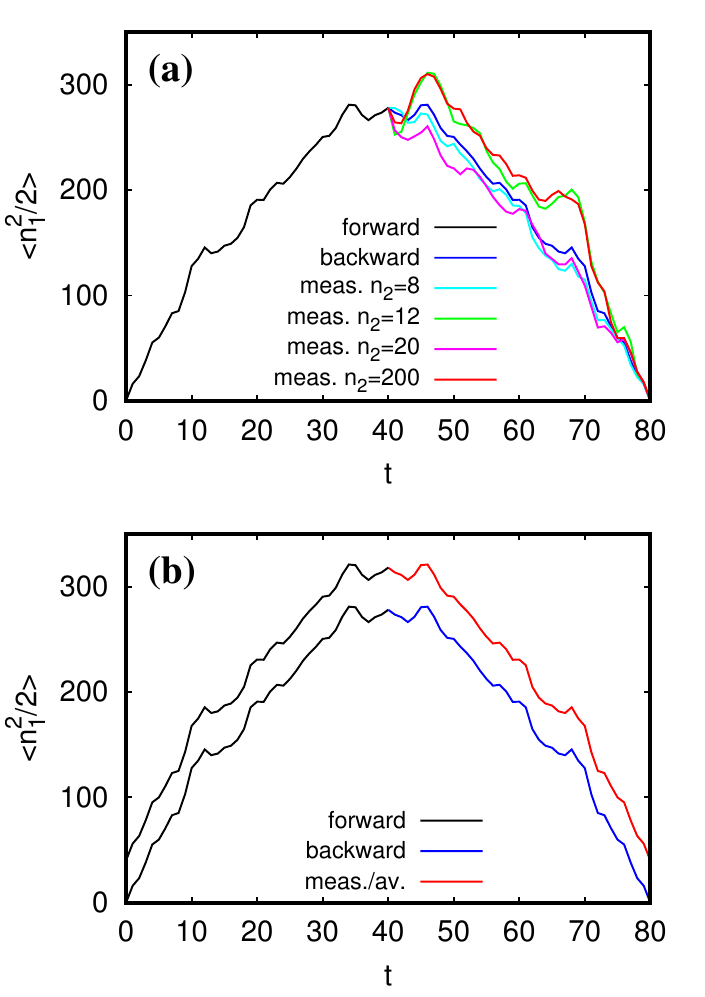}
\end{center}
\caption{\label{fig1} 
Time dependence of the average energy 
of the first particle $E_{1}(t) = \langle n_{1}^2/2\rangle$
for the initial state (\ref{eqinitialstate}) 
with time evolution given by the quantum Chirikov standard map (\ref{qmap})-(\ref{qmappair}).
The measurement of second particle and time reversal are performed
after $t_r =40$ quantum map (\ref{qmappair}) iterations.  
The black curves in both panels show the forward time evolution 
for $0\le t\le t_r$; the blue curves show the 
backward time evolution $t_r \le t \leq 2t_r=80$ 
with the exact time reversal without measurement (using $T=4\pi-\epsilon$). 
In panel (a) the curves of other colors show the backward time evolution
after measurement of the second particle at momentum states $n_2=8$
(cyan), 12 (green), 20 (magenta) , 200 (red).
In panel (b) the red curve shows the backward time evolution 
after the second particle measurement detection at $n_2$ and  
averaging over all possible measurement results of $n_2$ values 
(black and red curves are shifted up by 50 units for a better visibility;
red and blue curves coincide within numerical 
round-off errors ($\sim 10^{-13}$)). The system parameters are:
 $N=1024$, $N_H=N^2$ and $\hbar_{\rm eff}=\epsilon=5/8$, $K_{\rm eff}=5$, 
$k=K_{\rm eff}/\hbar_{\rm eff}=8$,
$T=4\pi \pm \epsilon$. We have verified that a further increase of $N$ 
to values of $2048$ and $4096$ provide identical results up to numerical 
round-off errors (provided the free propagation eigenphases are properly 
computed as explained in the text at the beginning of this section).
}
\end{figure}

\subsection{IIIA. EPR pairs in kicked rotator at moderate $\hbar_{\rm eff}$ values}
\label{subsec3a}

Here we present the results for a case with moderate effective value 
of the Planck constant.
As described above we use the values of parameter 
$T = 4\pi +\epsilon$ for forward
time propagation with $t_r$ quantum map iterations and $T=4\pi - \epsilon$
for next $t_r$ iterations corresponding to the time reversal. We remind that
since the phase shift $(4\pi) n^2/2$ is a multiple of $2\pi$ for all integer
values of the momentum $p=n$ the evolution is determined by an effective
Planck constant $\hbar_{\rm eff}=\epsilon$. Thus the effective 
classical chaos parameter 
is $K_{\rm eff} = k \epsilon = k \hbar_{\rm eff}$. The measurement is done 
for the second particle after $t_r$ iterations.
We consider the case of projective measurement in the momentum basis
$n_2$ of the second particle  performing the projection 
to a certain value of $n_2$ after $t_r$ iterations. After that the 
evolution of the first particle continues with $T=4\pi - \epsilon$
and $k \rightarrow -k$
for the next $t_r$ iterations. Without measurement the EPR wavefunction 
of two particles returns
exactly to its initial state due to exact time reversibility
of the quantum evolution. Also, in absence of entanglement of particles
the measurement of the second particle does not affect the reversibility
of the first particle which would exactly return to its initial state.
However, in presence of entanglement the measurement
of the second particle affects the time reversibility of the first 
particle in a nontrivial manner.

To illustrate the nontrivial features of
measurements on time reversal 
of chaotic EPR pairs we use typical 
system parameters with $K=k \epsilon = k\hbar_{\rm eff} =5$
and $k=8$ (thus $\hbar_{\rm eff} =5/8$). 
Such a value of $k=8$ is 
not very high being well accessible to the present experimental facilities
(see e.g. \cite{raizen,garreau,hoogerland}).

In this first part to characterize the quantum time evolution 
we compute the one-particle probability (of the first particle) as: 
$w(n_1,t) = \sum_{n_2}|\psi(n_1,n_2,t)|^2$,
(with the momentum wave function 
$\psi(n_1,n_2,t)=\langle n_1,n_2\ket{\psi(t)}$), 
and the one-particle energy (of the first particle):
$E_{1}(t) = \langle n_{1}^2/2\rangle=\sum_{n_1} (n_1^2/2)\,w(n_1,t)$.

As initial state we take an entangled EPR pair 
without any symmetry and with 
more or less arbitrary coefficients at two momentum values:
\begin{eqnarray}
\label{eqinitialstate}
\ket{\psi(t=0)}&=&\Bigl(
\ket{0}\otimes\ket{0}+
0.7\ket{0}\otimes\ket{1}+\\
\nonumber
&&\quad 0.3\ket{1}\otimes\ket{0}-
2\ket{1}\otimes\ket{1}\Bigr)/\sqrt{5.58} \; ,
\end{eqnarray}
where $\ket{n_1}\otimes\ket{n_2}$ represents the momentum basis states.
Thus initially both particles are distributed over
momentum states at $n_{1,2}$ being $0$ or $1$.

This state can be rewritten in the Schmidt decomposition \cite{schmidt} as~:
\begin{equation}
\label{eqschmidt}
\ket{\psi(t=0)}=\sum_{i=1,2} \alpha_i \ket{u_i}\otimes\ket{v_i}
\end{equation}
with
\begin{eqnarray}
\label{eqschmidt2}
\nonumber
\alpha_1&=&0.8973\quad,\quad\alpha_2=0.4414,\\
\nonumber
\ket{u_1}&=& 0.3440\ket{0}-0.9390\ket{1},\\
\nonumber
\ket{u_2}&=& 0.9390\ket{0}+0.3440\ket{1},\\
\nonumber
\ket{v_1}&=& 0.0294\ket{0}+0.9996\ket{1},\\
\ket{v_2}&=& 0.9996\ket{0}-0.0294\ket{1},
\end{eqnarray}
The entropy of entanglement of this initial state is:
\begin{equation}
\label{eqentropy_log2}
S=-\sum_i \alpha_i^2\,\log_2(\alpha_i^2)=0.7114 \; .
\end{equation}

\begin{figure}[h]
\begin{center}
\includegraphics[width=0.45\textwidth]{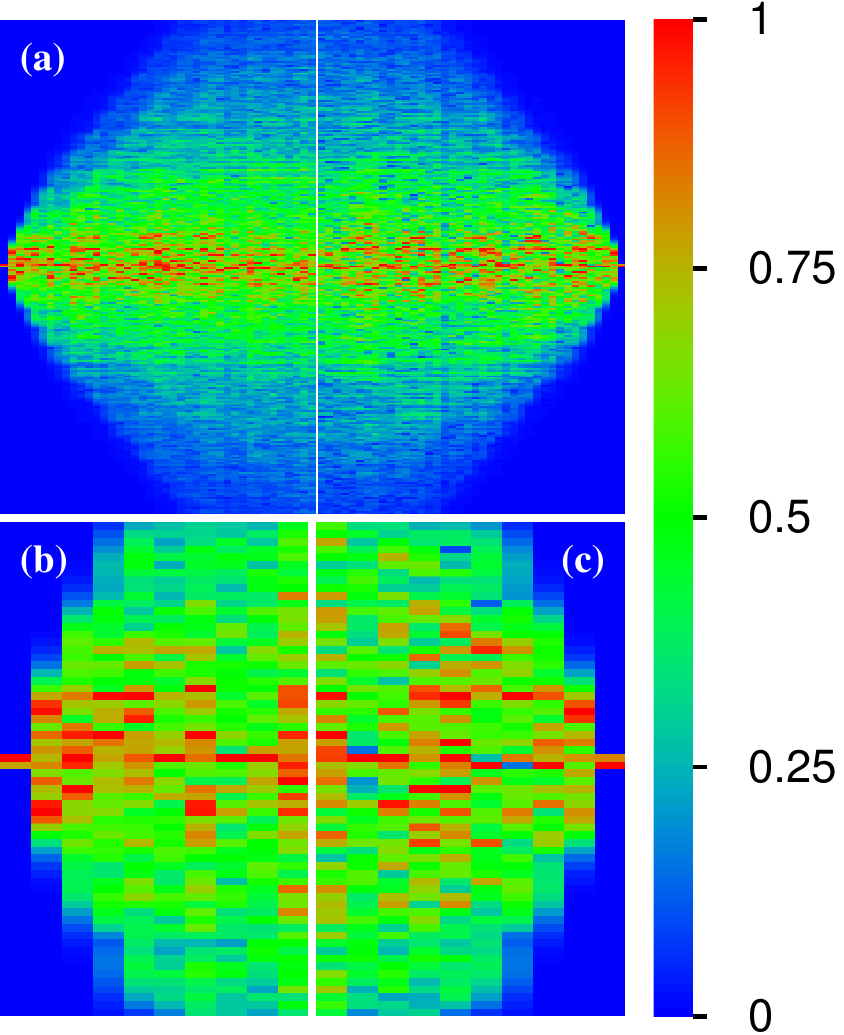}
\end{center}
\caption{\label{fig2} Panel (a) shows the time evolution 
of probability of the first particle $w(n_1,t)$ (color density plot) 
for parameters of Fig.~\ref{fig1} and $-128\le n_1<128$ 
($y$-axis), $0\le t\le 80$ ($x$-axis), $t_r=40$. 
The measurement and time reversal are done after $t_r$ map (\ref{qmappair}) iterations
with the second particle detected at the momentum value $n_2=12$.
The thin white vertical 
line marks the time $t_r=40$ of measurement   
and the beginning of backward iterations. 
The numbers of the color bar correspond to 
$[w(n_1,t)/w_{\rm max}(t)]^{1/4}$
with $w_{\rm max}(t)=\max_{n_1} w(n_1,t)$ being the density maximum 
at a given  value of $t$. 
Panels (b) and (c) provide a zoom 
for $-32\le n_1<32$ (both panels) 
and $0\le t<10$ (b) or $70<t\le 80$ (c). 
}
\end{figure}

\begin{figure}[h]
\begin{center}
\includegraphics[width=0.45\textwidth]{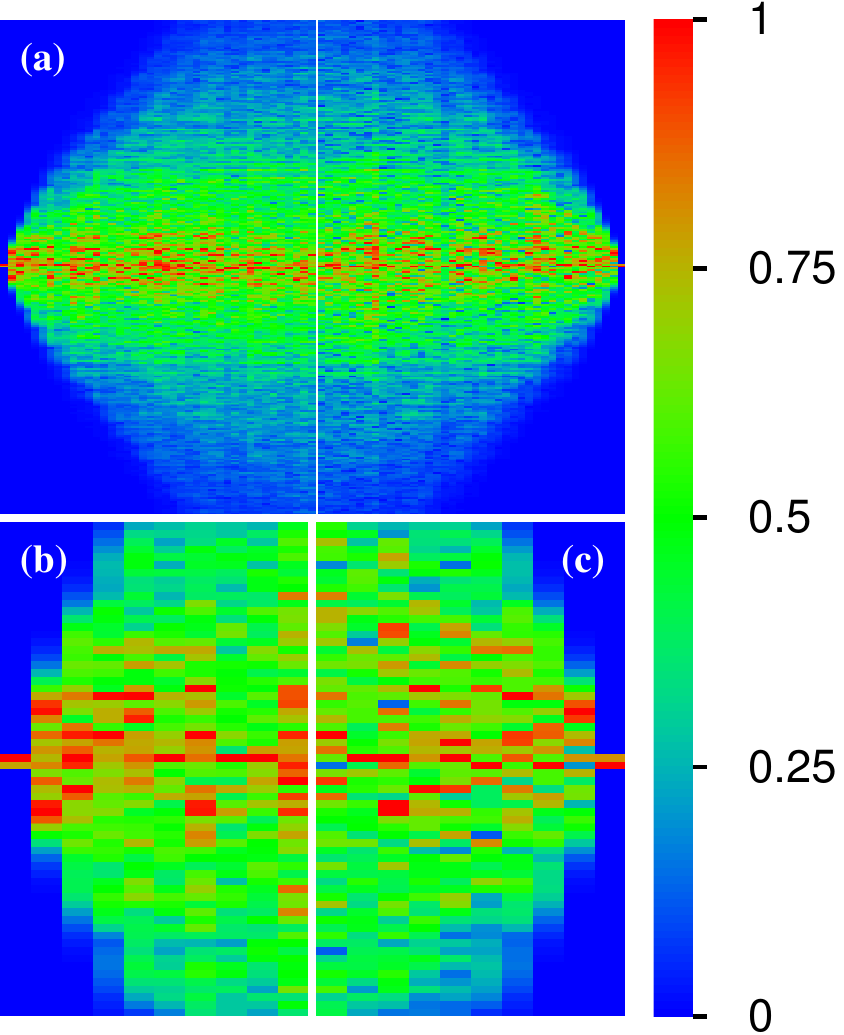}
\end{center}
\caption{\label{fig3} Same as Fig.~\ref{fig2} but for  
the second particle  measured at $n_2=200$. 
}
\end{figure}

The time evolution of energy of the first particle
$E_{1}(t) = \langle n_{1}^2/2\rangle$ is shown in Fig.~\ref{fig1}.
At short initial times $ t \leq 12$ we have an
approximately diffusive energy growth
$E_1(t) \approx Dt/2 \approx 16 t$ with the diffusion coefficient
$D \approx k^2/2 \approx 32$. For times $12 < t  \leq t_r=40$ 
the energy growth continues but its rate decreases due to the quantum interference effects
being similar to the Anderson localization \cite{chi1981,chi1988,dls1983}). 
After $t_r=40$ iterations of forward propagation in time,
the projective measurement is performed for the second particle
and the evolution of the first particle is reversed in time
with a replacement $T= 4\pi + \epsilon \rightarrow 4\pi - \epsilon$
and $k \rightarrow -k$ (effective backward propagation in time).

In the top panel Fig.~\ref{fig1}(a) we show the energy
$E_1(t)$ dependence on time for forward $0 \leq t \leq t_r$ 
and backward evolution $t_r < t \leq 2t_r$
for different results of projective measurement, 
done after $t_r=40$ iterations,
giving (projecting) the second particle at
different momentum states chosen as $n_2=8; 12; 20; 200$.
For each measured $n_2$ value we have a different curve 
$E_1(t)$ of the time reversal or backward branch $t_r< t \leq 2t_r$
being different from the forward branch  $0 \leq t \leq t_r$.
However, all curves at different $n_2$ measured values
have an energy decrease with time and approximately 
return to the initial energy value.
Of course, in absence of measurements there is the exact time
reversal of evolution and energy $E_1(t)$ is exactly symmetric with 
respect to the moment of time reversal and returns exactly to the initial
value as it is shown in Fig.~\ref{fig1}(b) (blue curve).
In the same panel we also present the result of
backward evolution of $E_1(t)$ averaged over 
all projective measurements of second particle 
found at all possible momentum values $n_2$ (Fig.~\ref{fig1}(b) red curve).
The red and blue curves coincide up to 
numerical round-off errors being on a level of $10^{-13}$.
Such an exact coincidence of time reversal behavior 
without measurements and with  averaging over all
possible measurement results 
can be understood from the Feynman path integral formulation
of quantum mechanics \cite{feynman}. In this Feynman interpretation
a specific projective measurement of second particle at  $n_2$ value
selects a specific entangled path of first particle
which returns approximately to its initial state at $t=0$.

Examples of the time evolution of the probability distribution
of the first particle $w(n_1,t)$ are shown in Figs.~\ref{fig2},~\ref{fig3}
for two cases of projective measurements of the second particle
at momentum values $n_2 = 12, 200$ respectively
(see also Appendix Fig.~\ref{figA1}).
The measurement is done after $t_r=40$ quantum map (\ref{qmappair}) iterations.
For $0 \leq t \leq t_r$ the values of  $w(n_1,t)$
are obtained by averaging the two-particle density over the second particle,
after measurement of second particle  $w(n_1,t)$
represents the probability distribution 
over momentum states of the remaining first particle.
The results of Figs.~\ref{fig2},~\ref{fig3}
show a diffusive type spreading of probability $w(n_1,t)$
during the time range $0 \leq t \leq t_r$.
This corresponds to a diffusion produced 
by the underlined classical chaotic dynamics
(quantum corrections give a certain  reduction of 
the diffusion rate as discussed above
and in \cite{chi1981,chi1988}). After the projective measurement of the 
second particle and the time reversal of the propagation 
an inverse diffusion process takes 
place where the 
probability   $w(n_1,t)$ returns approximately
to the initial state of the first particle. 
This corresponds to a specific Feynman path selected 
by the projective measurement of the second particle
immediately after the time $t_r$. 
Some snapshots of the probability distribution $w(n_1,t)$ 
corresponding to different results of measurements 
with different $n_2$ values are shown for
specific time moments in Appendix Fig.~\ref{figA2}.

\begin{figure}[h]
\begin{center}
\includegraphics[width=0.45\textwidth]{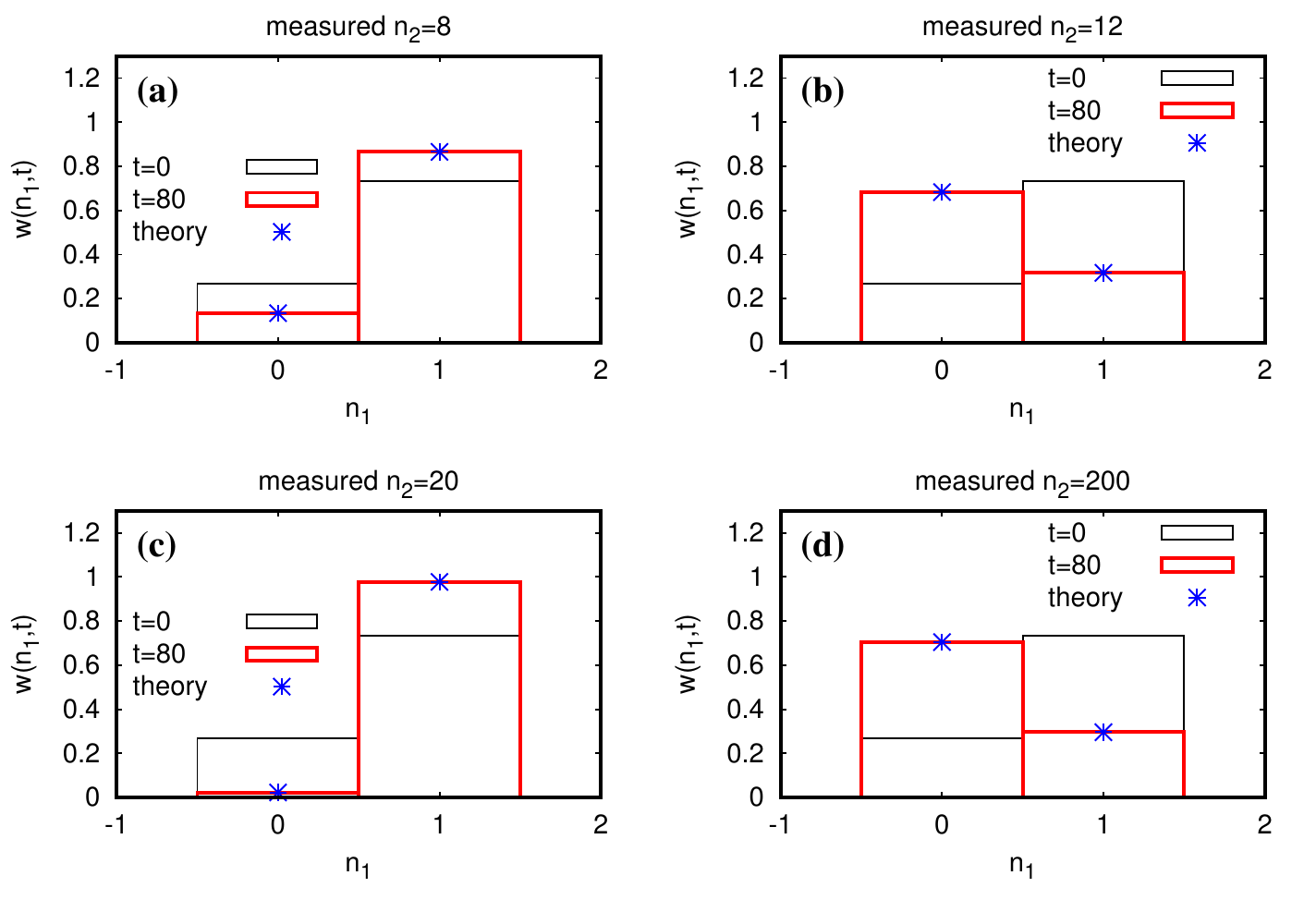}
\end{center}
\caption{\label{fig4} Probabilities $w(n_1,t)$ of the first particle
at initial time $t=0$ (black lines)
and final time $t=2t_r=80$ (red lines) for the cases when the second particle is measured 
at $n_2=8$ (a), $n_2=12$ (b), $n_2=20$ (c) and $n_2=200$ (d) after 
the time reversal at $t_r=40$; system parameters are as in Fig.~\ref{fig1}. 
The blue stars provide for each case the semi-analytical theoretical prediction for the 
final density at $t=2t_r=80$ obtained from the Schmidt decomposition (see text).
}
\end{figure}

The probability distribution $w(n_1,t=2t_r)$ 
at the reversal time $t=2t_r$ is shown in Fig.~\ref{fig4}.
All probability is located at momentum states $n_1=0,\,1$ 
corresponding to the states populated by the first 
particle in its initial entangled state (\ref{eqinitialstate})
(probability outside of these states
is on a level of numerical round-off errors $10^{-13}$).
However, the values of the two return probabilities $w(n_1,t=2t_r)$
are affected by the measurement of the second particle and they are 
rather different from their values of 
the initial state (\ref{eqinitialstate}).

The values of $w(n_1,t=2t_r)$ can also be computed from a ``theoretical'' 
state $\ket{u_{\rm \,th}}$ for the first particle obtained 
from the assumption that 
only the second particle is following the time evolution while 
the first particle remains fixed and measuring the second particle 
at $t=t_r$. The theoretical state is then given by 
$\ket{u_{\rm \,th}}=\alpha_1\,C_1\ket{u_1}+\alpha_2\,C_2\ket{u_2}$ 
where (for $j=1,2$) $\alpha_j$ and $\ket{u_j}$ are given by Eq. 
(\ref{eqschmidt2}) used for 
the Schmidt decomposition of the initial state (\ref{eqschmidt}). 
The coefficients $C_j$ are obtained from the measurement procedure as 
$C_j=C_g\,\langle n_2\ket{v_j(t_r)}=C_g\,v_j(n_2,t_r)$ 
where $\ket{v_j(t_r)}$ are the second-particle states at the 
moment of measurement (after $t_r$ iterations) 
and $C_g$ is the global normalization constant 
of the theoretical state $\ket{u_{\rm \,th}}$. 
The blue stars in Fig.~\ref{fig4} show the values obtained from this 
theoretical state which coincide numerically (up to usual round-off errors) 
with the values of $w(n_1,t=2t_r)$ (also for the cases $n_1$ being different 
from 0 or 1 where are simple $w(n_1,t=2t_r)=0$).

The reason is that the Schmidt coefficients $\alpha_i$ remain 
unchanged during the forward propagation in time till 
the moment $t=t_r$ while the Schmidt vectors $u_1(n_1,t), v_1(n_2,t)$ and 
$u_2(n_1,t), v_2(n_2,t)$ evolve as one-particle wavefunctions
computed numerically from (\ref{qmap})
with the initial condition (\ref{eqschmidt}),(\ref{eqschmidt2}).
The measurement of the second particle, detected at momentum state $n_2$
at $t=t_r$, gives the above coefficients $C_1$ and $C_2$ and 
after the measurement the wavefunction of the first particle is:
$\psi(n_1,t) = \alpha_1 C_1 u_1(n_1,t) + \alpha_2 C_2 u_2(n_1,t)$.
During the backward evolution the components $ u_1(n_1,t)$ and
$ u_2(n_1,t)$ return to their initial values providing exactly 
the theoretical state given above. 
Hence, the probability to find the first particle
at $n_1=0$ is $w(n_1=0)=(0.3440 \alpha_1 C_1 + 0.9390 \alpha_2 C_2)^2$
and at $n_1=1$ it is 
$w(n_1=1) = (-0.9390 \alpha_1 C_1 + 0.3440 \alpha_2 C_2)^2$.
As it is shown in Fig.~\ref{fig4}
the results of this semi-analytical theory
reproduce the numerically obtained probabilities (up to 
usual round-off errors).

\begin{figure}[h]
\begin{center}
\includegraphics[width=0.45\textwidth]{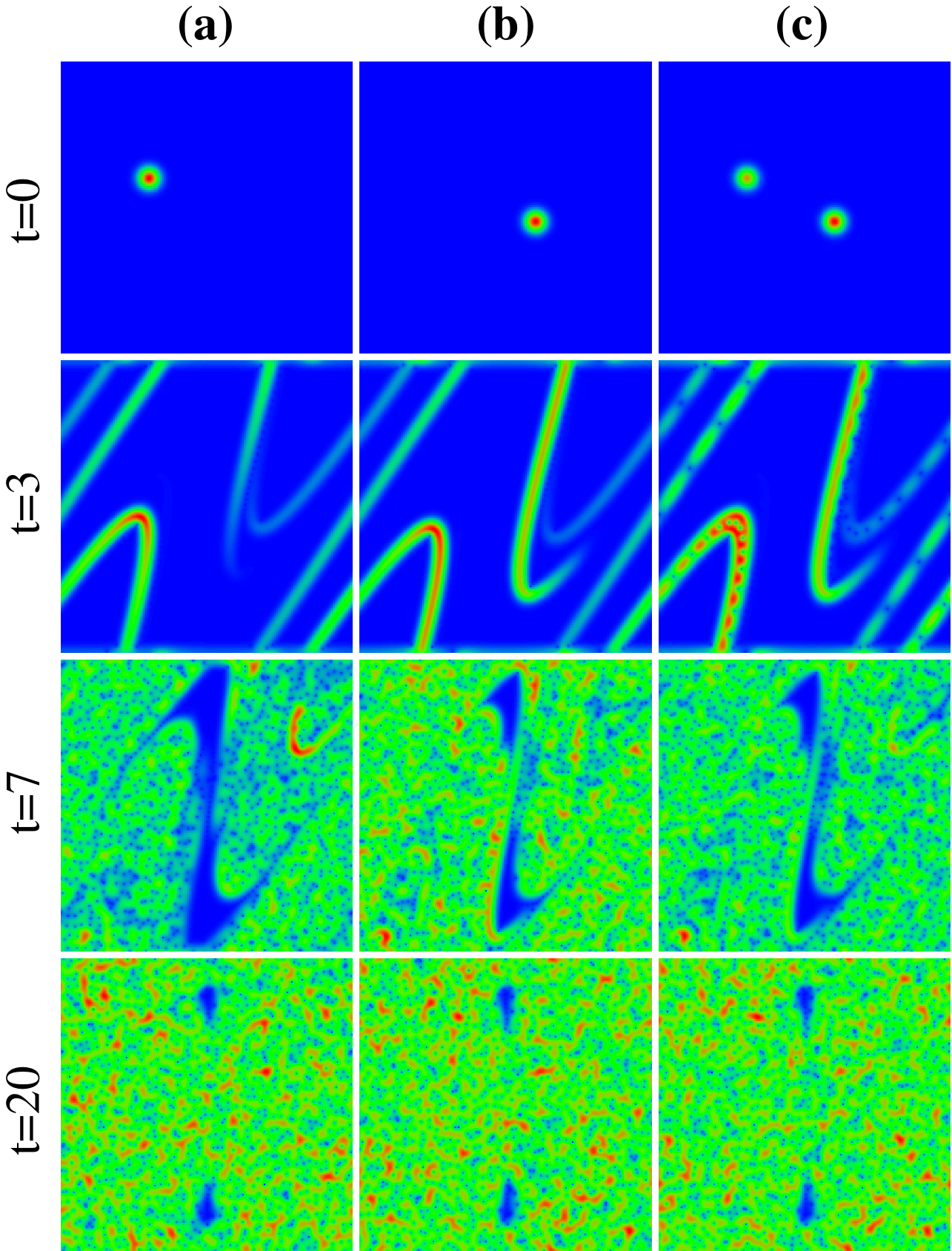}
\end{center}
\caption{\label{fig5} Density plots of Husimi functions of the 
first particle at  certain time moments  $t$. 
Columns (a) and (b) show the Husimi functions 
of the time dependent Schmidt components $\ket{u_1(t)}$ (a) and 
$\ket{u_2(t)}$ (b) of the first particle after 
$t=0,\,3,\,7,\,20$ iterations 
with the horizontal axis corresponding to $x_1\in[0,\,2\pi[$ and the 
vertical axis corresponding to $p_1 \in[-\pi,\,\pi[$ 
(or $n_1\in[-N/2,\,N/2[$ with  $p_1= \hbar_{\rm eff} n_1$
for $N=2^{10}$).
The initial conditions $\ket{u_j(t=0)}$ (for $j=1,2$) 
are Gaussian coherent states (see  (\ref{eqcoherent}) and text).
The initial Schmidt components $\ket{v_j(t=0)}$ of 
the second particle are also Gaussian coherent states of the same type with 
close positions (see text). 
Column (c)  shows the Husimi functions of the first particle after the 
second particle being 
measured at $n_2=8$  after $t=t_r=20$ iterations and 
followed by time reversal. The iteration times for the right column are 
$t_{(c)}=40-t$ with $t$ being the time values shown in the figure 
for each row, i.e. $t_{(c)}=40,\,37,\,33,\,20$ (top to bottom).
The color bar is the same as in Figs.~\ref{fig2} and \ref{fig3} where the 
numbers correspond to $[H(x,p)/H_{\rm max}]^{1/4}$ with 
$H(x,p)$ being the Husimi function 
of the first particle. System parameters are: 
$\epsilon=\hbar_{\rm eff} = 2\pi/N = \pi \times 2^{-9}$,
$K_{\rm eff}=k \hbar_{\rm eff}=5$, $T=4\pi \pm \epsilon$.
The pixel resolution of each panel corresponds to $P\times P$ pixels with 
$P=8\sqrt{N}=2^8$.
}
\end{figure}

\begin{figure}[h]
\begin{center}
\includegraphics[width=0.45\textwidth]{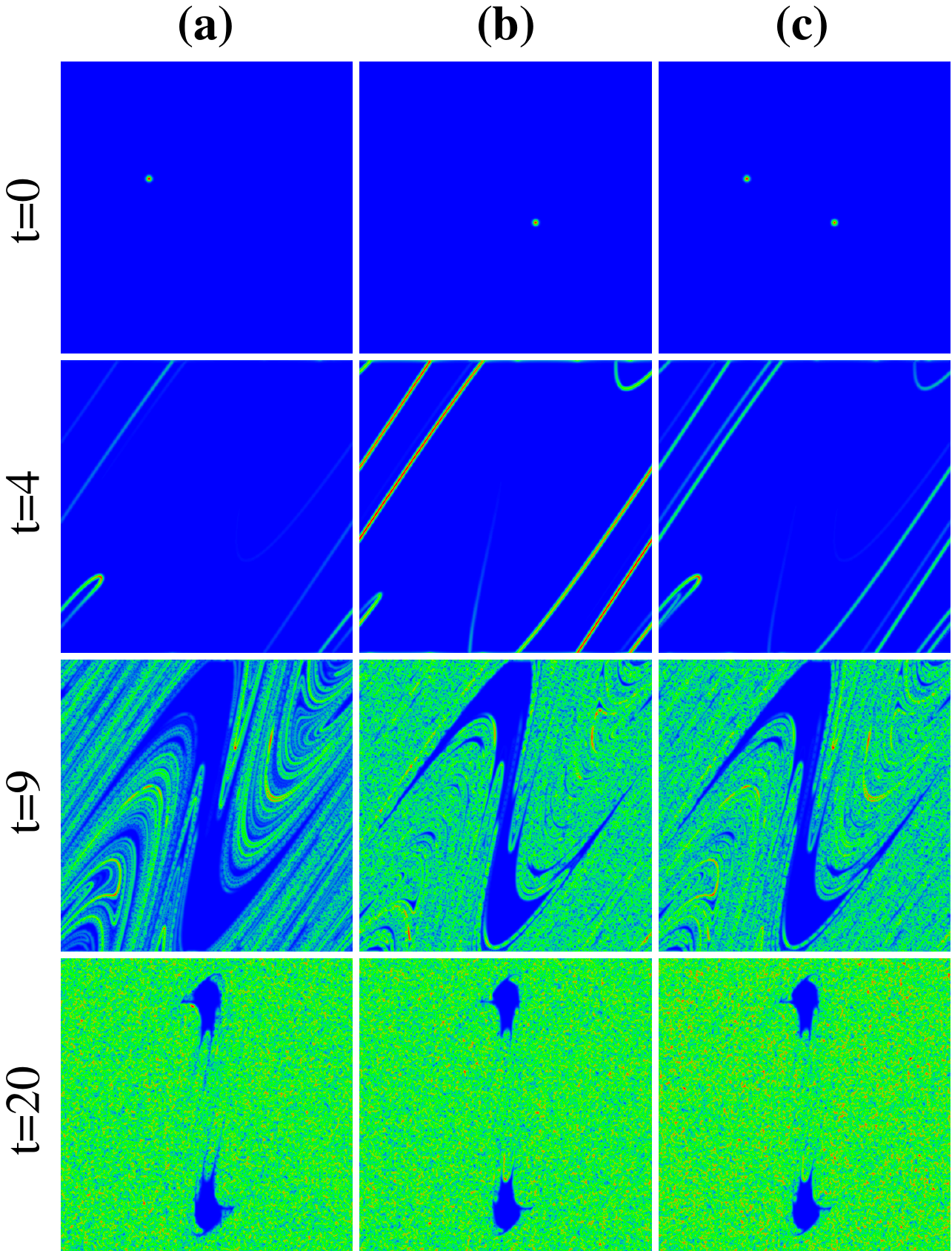}
\end{center}
\caption{\label{fig6} Same as Fig.~\ref{fig5} 
but for $N=2^{14}$ and modified 
iterations times $t=0,\,4,\,9,\,20$ (a), (b) 
or $t_{(c)}=40-t=40,\,36,\,31,\,20$ (c) (top to bottom);
$\epsilon = \hbar_{\rm eff}=2\pi/N= \pi \times 2^{-13}$, 
$K_{\rm eff}= k \hbar_{\rm eff}=5$.
The number of pixels per direction is $P=8\sqrt{N}=2^{10}$.
}
\end{figure}

\subsection{IIIB. EPR pairs in kicked rotator at small $\hbar_{\rm eff}$ values}
\label{subsec3b}

We also study the behavior of chaotic EPR pair in a deep
semiclassical limit. For this we consider the quantum evolution on a torus
of size $N$ with $\epsilon = \hbar_{\rm eff} = 2\pi/N$ and 
periodic conditions for the wavefunction in a momentum space
$-\pi \leq p =\hbar_{\rm eff} n <\pi$. Such a time evolution of 
quantum maps on a quantum torus
has been already studied in detail for the 
one-particle case (see e.g. \cite{chi1988,qcfrahm}).
As above, we choose the effective chaos parameter 
$K_{\rm eff} =  \hbar_{\rm eff} k =5$
such that the phase space of classical dynamics is mainly chaotic
with only small integrable islands embedded in the chaotic component
(the measure of these islands is approximately 2\% \cite{chirikov1979}).

The initial state of the EPR pair is chosen as an entangled state 
given as the Schmidt decomposition of two pairs of coherent 
Gaussian states and with equal coefficients $\alpha_1=\alpha_2=1/\sqrt{2}$. 
More precisely, the initial components of the first particle are
\begin{eqnarray}
\label{eqcoherent}
&&\ket{u_j(t=0)} = \ket{(x_0^{(j)},p_0^{(j)})_{\rm coh.}}\\
&&\qquad = C\sum_{p} \exp[-G(p-p_0^{(j)})^2 
\nonumber
-i p \,x_0^{(j)}/\hbar_{\rm eff}]\ket{p} 
\end{eqnarray}
with classical positions of the coherent wave packet at 
$x_0^{(1)}=0.3\times (2\pi)$, $p_0^{(1)}=0.1\times (2\pi)$ and 
$x_0^{(2)}=0.6\times (2\pi)$, $p_0^{(2)}=-0.05\times (2\pi)$. 
(Note that in (\ref{eqcoherent}) the $p$-sum runs over the values 
$p=2\pi n/N$ with $n=0,\,\ldots,\,N-1$.)
The parameter $G$ is related to the width 
$\Delta p=\langle (\hat p-p_0^{(j)})^2\rangle^{1/2}$ of 
the wave packet in momentum space by $G=1/(4\Delta p^2)$ 
and $C$ is the normalization constant. 
Here we choose $G=1/(2\hbar_{\rm eff})=N/(4\pi)$ such that 
$\Delta p=\sqrt{\pi/N}$ and $\Delta x=\hbar_{\rm eff}/(2\Delta p)=
\sqrt{\pi/N}$ are identical (note that 
here for both variables $x$ and $p$ the system size is $2\pi$). 

The initial Schmidt components $\ket{u_j(t=0)}$ of 
the second particle are also Gaussian coherent states of the same type 
(\ref{eqcoherent}) 
with  close initial positions 
$\ket{v_j(t=0)}=\ket{(x_0^{(j)}+\delta x,p_0^{(j)}+\delta p)_{\rm coh.}}$
shifted by  $\delta x=\delta p=2\pi/\sqrt{N}=2\sqrt{\pi}\Delta p$. 

We describe the probability distribution of the first particle 
components (\ref{eqcoherent})
in the phase space $(x_1,p_1)$ by the Husimi function 
defined as $H(x,p)=|\langle (x,p)_{\rm coh.}\ket{u}|^2$ 
where $\ket{u}$ is the first-particle state for which the 
Husimi function is computed and $\ket{(x,p)_{\rm coh.}}$ is the 
coherent state (\ref{eqcoherent}) with the above choice of 
$G=1/(2\hbar_{\rm eff})$ at classical positions $(x,p)$. 
The Husimi function corresponds to a smoothing of the 
Wigner function over the above values of $\Delta x$, $\Delta p$ 
with $\Delta x \Delta p = \hbar_{\rm eff}/2$ 
(see e.g. \cite{qcfrahm,husimi} for a description of Husimi functions). 
We have numerically computed the Husimi function using an efficient algorithm 
based on FFT and for a grid size $(\delta x/8)\times (\delta p/8)$ (with the 
above values of $\delta x$ and $\delta p$).

The Husimi functions of the first particle Schmidt components $u_1, u_2$
are shown in Fig.~\ref{fig5} at different iteration times 
$t=0,\,3,\,7,\,20$ (in columns (a) and (b)) for
$\hbar_{\rm eff}=2\pi/N$ and $N=1024$. 
We see that due to the underlying classical chaos the initial coherent state
spreads very rapidly over the whole available phase space,
except the domain of integrable islands which can be occupied
only after very long tunneling times. In column (c) of Fig.~\ref{fig5}
we show the backward time evolution of the first particle wave packet
obtained from the measurement of the second particle 
at the momentum state $n_2=8$ performed after $t_r=20$ quantum map 
(\ref{qmappair}) iterations. 
We see that the backward evolution of the first particle has 
different Husimi distributions at different return time moments.
However, at $t=2t_r=40$ the first particle returns to its initial coherent
states of Schmidt components at $t=0$ but with different weights.
In Fig.~\ref{fig6} we show also a similar time evolution with 
measurement and time
reversal for the smaller value $\hbar_{\rm eff} =2\pi/N$, $N=10^{14}$ and 
the measured state $n_2=8$.
As in Fig.~\ref{fig5} there is an exponentially rapid spreading of initial 
coherent wave packets which after measurement returns to the initial 
two coherent states but with different weights.

\begin{figure}[h!]
\begin{center}
\includegraphics[width=0.45\textwidth]{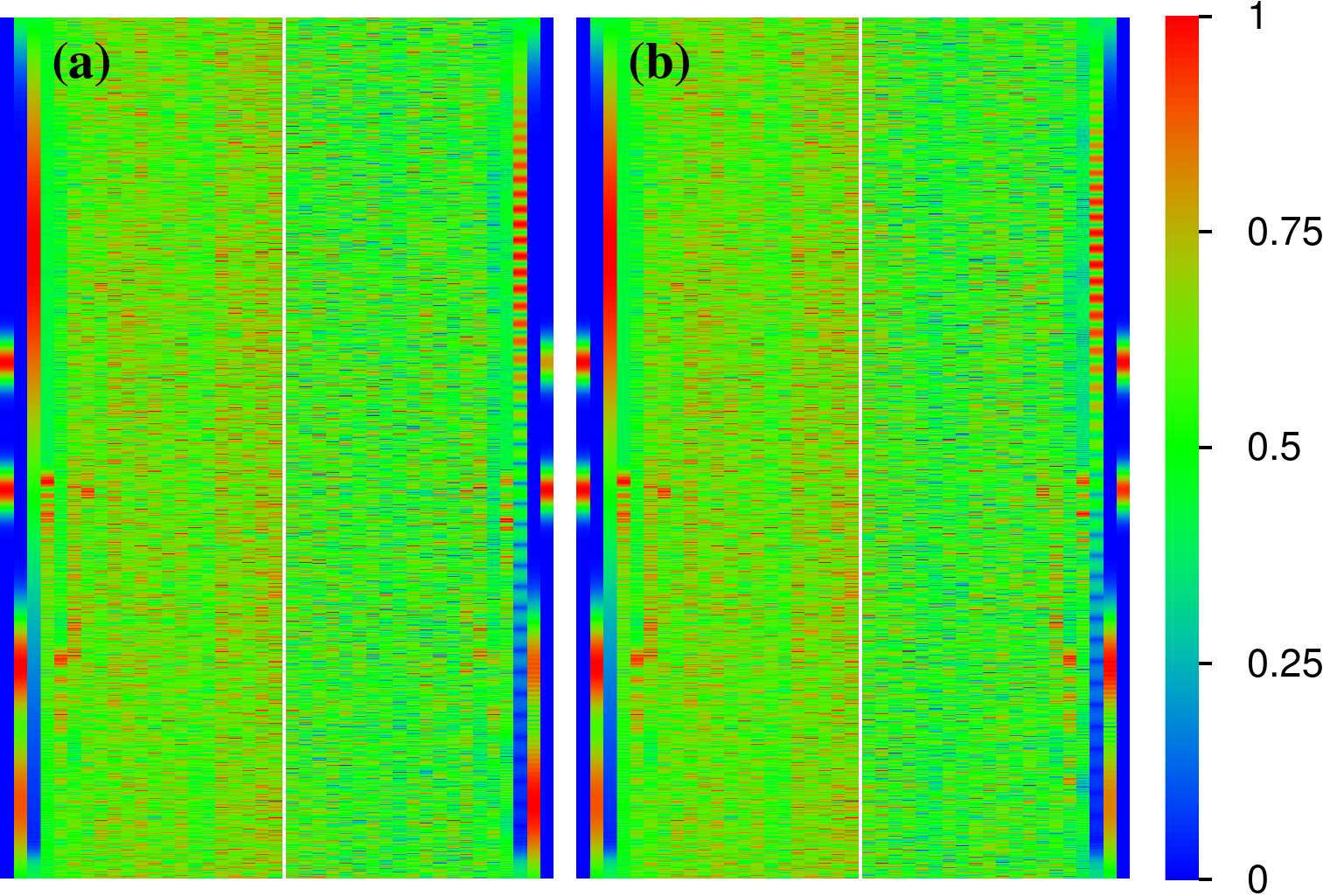}
\end{center}
\caption{\label{fig7} Panels show color 
density plot of $w(n_1,t)$ for the first particle  time evolution and 
parameters of Fig.~\ref{fig5} with $-512\le n_1<512$ 
($y$-axis) and $0\le t\le 40$ ($x$-axis). The time values 
for $t\le t_r=20$ correspond to the exact forward iteration 
and for backward iterations $ t_r < t \leq 2t_r =40$,
the measurement, done
after $t_r$ iterations, detects the second particle at $n_2=8$ (a) or $n_2=200$ (b). 
The thin vertical white 
line marks the time position of measurement  $t_r=20$ 
and the beginning of the time reversal backward iterations. The  
color bar has the same meaning as in Fig.~\ref{fig2}.
}
\end{figure}

In Fig.~\ref{fig7} we show the time evolution of 
probability distribution $w(n_1,t)$ of first particle over
momentum states $n_1$. For $0 \leq t \leq t_r$ this probability is
averaged ob the second particle. As in Fig.~\ref{fig5}
the projective measurement
is done after $t_r=20$ quantum map (\ref{qmappair})  iterations
with the second particle detected at $n_2=8$.
We see that the backward probability distribution $w(n_1,t)$
at $t>t_r$ is different from the forward one.
However, at the return moment $t=2t_r$ we still
have two coherent wave packets for first particle
which have the same shape as at the initial state
but with different coefficients. 

\begin{figure}[h]
\begin{center}
\includegraphics[width=0.45\textwidth]{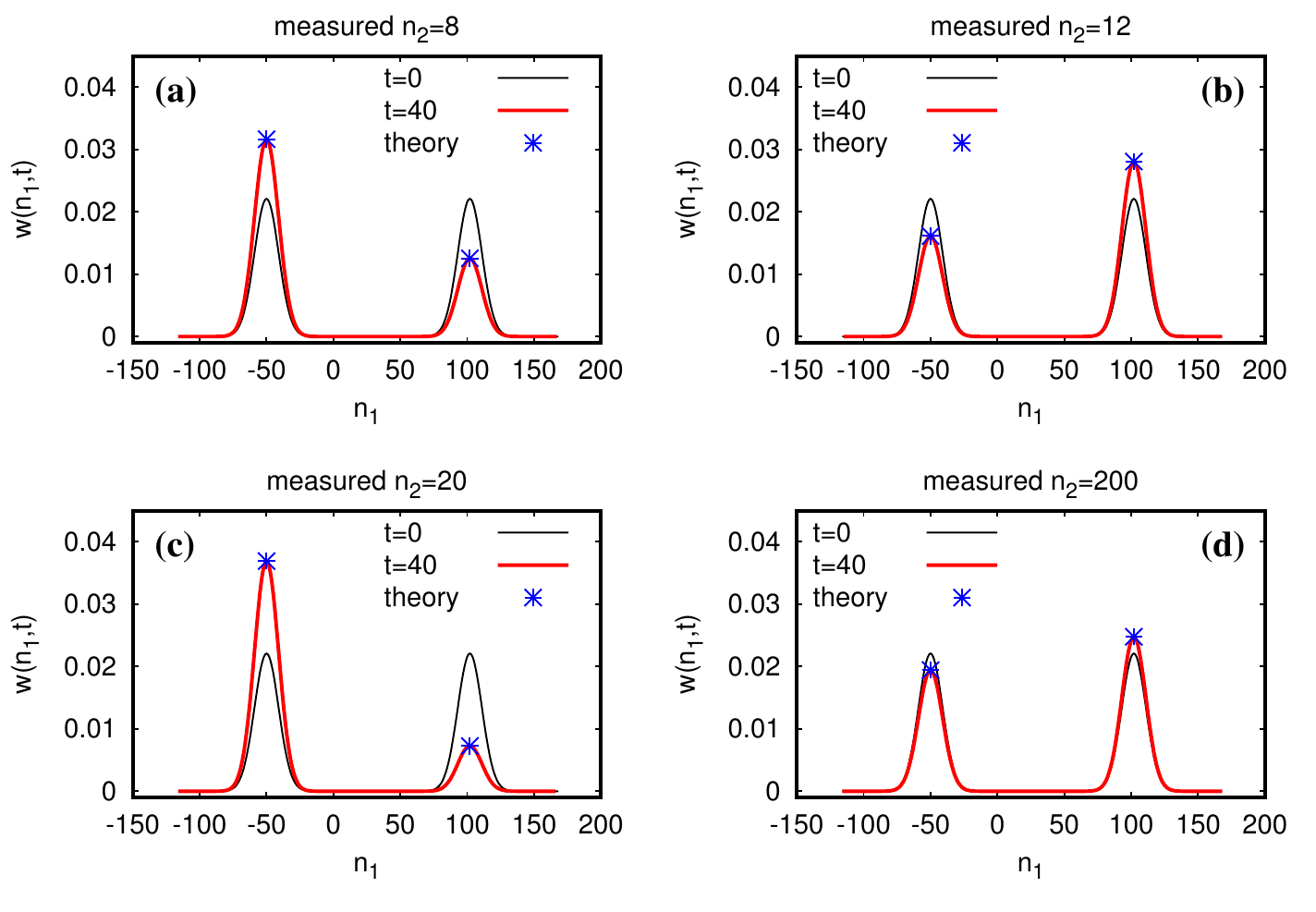}
\end{center}
\caption{\label{fig8} Densities $w(n_1,t)$ of the first particle
for the case of Fig.~\ref{fig5} are
shown at initial $t=0$ (black curves)
and final $t=2t_r=40$ (red curves) time moments;
 the second particle  is measured 
at $n_2=8$ (a), $n_2=12$ (b), $n_2=20$ (c) and $n_2=200$ (d). 
The blue stars provide for each case the maximum values of the 
density of the theoretical state obtained from the Schmidt decomposition 
(see text) and predicting the final density at $t=40$.
The theoretical density curves are identical to the red curves within 
numerical precision $\sim 10^{-13}$ but only their values at the two maximum 
positions are shown for a better visibility.
}
\end{figure}

We also show the initial $t=0$
and final $t=2t_r=40$ probability distributions of the first particle
in Fig.~\ref{fig8} for different results of measurements
of the second particle detected at $n_2=8, 12, 20, 200$. 
The weights of each coherent state at $t=2t_r$ 
are determined from the Schmidt components of a theoretical state 
constructed in the same way as in the case of Fig.~\ref{fig4}.
The density of the theoretical state coincides with the final density 
at $t=2t_r$ up to usual numerical round-off errors (only the maximum of 
each theoretical state is shown in Fig.~\ref{fig8} by a blue star).

Similar to the case of Fig.~\ref{fig7}
with measured $n_2=8$ we show the time evolution $w(n_1,t)$
for other measured values $n_2 = 12, 20$ in Appendix Fig.~\ref{figA3}.
For comparison we show in Appendix Fig.~\ref{figA4}
also the case of exact time reversal without measurements
(i.e. with average over all measured $n_2$ values):
here the distribution $w(n_1,t)$ is exactly symmetric with respect to
time reversal at the moment $t=t_r=20$.

\subsection{IIIC. EPR pairs of cold atoms in a kicked optical lattice}
\label{subsec3c}

Above we studied the properties of measurements and time reversal of EPR pairs
in the regime of kicked rotator when the evolution takes place on a ring 
of size $2\pi$. However, the experiments with cold atoms 
in a kicked optical lattice \cite{raizen,garreau,hoogerland}
correspond to the situation when an EPR pair propagates on the 
infinite $x$ axis containing many periods of size $2\pi$. Due to 
the periodicity of potential the wavefunction of each
particle is characterized by a quasimomentum  with irrational values
$\beta$ (with $p=n+\beta$) that reduce the probability of the 
single atom time reversal as discussed in detail in \cite{martin}.
Thus to model this experimental setup we consider the EPR propagation on an 
$x$ interval of size $2\pi L$ containing $L$ periods $2\pi$ of the 
optical lattice. We use periodic boundary conditions in $x$
but during the time evolution the wave packet 
is not reaching the boundaries such that the 
boundary conditions are not important.
In this case the free propagation of a particle between kicks
is given by the same unitary operator
as in (\ref{qmap}) but now in numerical simulations
the momentum takes discrete values
$p=m/L$ with integers $m=-N/2,\,\ldots,\,, N/2-1$
and $N=L N_r$ where $L$ gives the number of different
quasimomentum values $\beta$ and $N_r$ gives the number of integer
values of momentum $p$.
The integer $p$ values corresponds to the rotator case.
The kick operator remains the same as in (\ref{qmap}) but the position 
operator now takes the discrete values $x=2\pi m L/N$ ($m$ having the 
same integer values as above) corresponding to the interval $[-\pi L,\pi L[$. 
As in the previous Sections the numerical simulations are done 
with the propagation of the full wavefunction using its Schmidt components.
This allows to reach very high $N$ and $L$ values required to
eliminate boundary effects. 
As it was shown in previous Sections this
computational method gives the same results
as  the full wavefunction propagation with 2D FFT
(up to numerical precision).
We use as  maximal values $N=2^{22}$  with $L=2^{14}$, $N_r=2^8$.

\begin{figure}[h!]
\begin{center}
\includegraphics[width=0.45\textwidth]{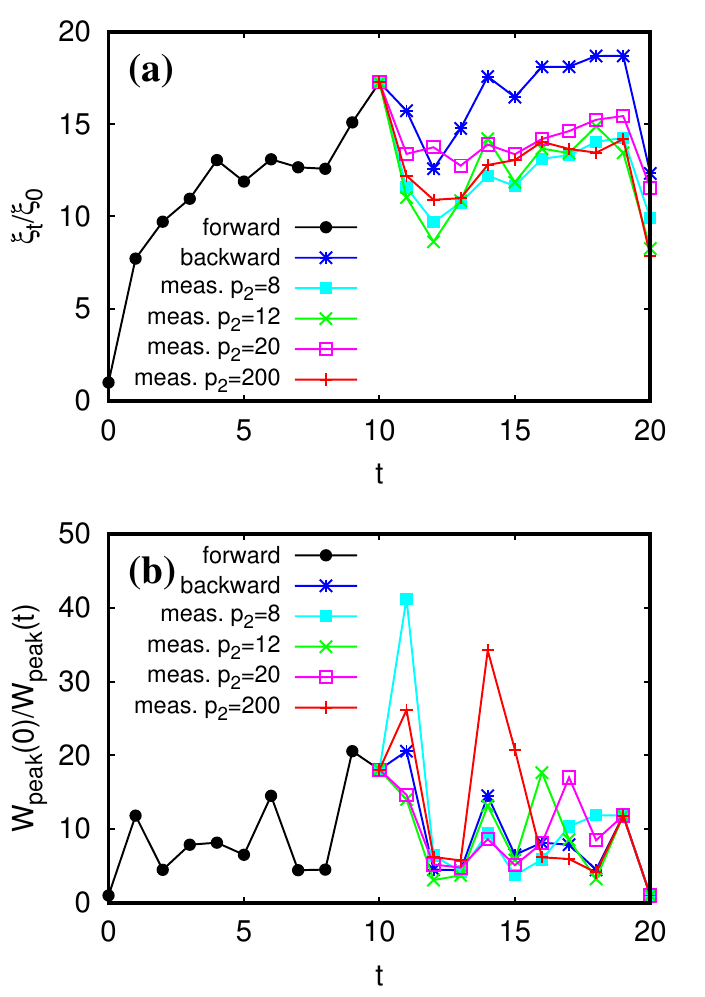}
\end{center}
\caption{\label{fig9} Time dependence of rescaled IPR
$\xi_t/\xi_0$ (a) and peak probability (b) of first particle
for a chaotic EPR pair in a kicked optical lattice:
the time forward evolution is marked by black points;
full backward evolution is marked by blue stars
(time reversal is done at $t=t_r=10$ without measurement),
backward evolution with measurement of the moment $p_2$ of the 
second particle done at $t_r$ is shown by different 
color symbols for different measured $p_2$ values
with $p_2=8$ (cyan full squares), $p_2=12$ (green crosses),
$p_2=20$ (magenta open squares), $p_2=200$ (red pluses).
System parameters are  $\hbar_{\rm eff}=\epsilon=5/8$, 
$K_{\rm eff}=5$, $k=K_{\rm eff}/\hbar_{\rm eff}=8$,
$T=4\pi \pm \epsilon$ (as in Fig.~\ref{fig1})
and $N=L N_r =2^{22}$, $L=2^{14}$, $N_r=2^8$.
The initial state of the EPR pair is described in the text.
}
\end{figure}

\begin{figure}[h!]
\begin{center}
\includegraphics[width=0.45\textwidth]{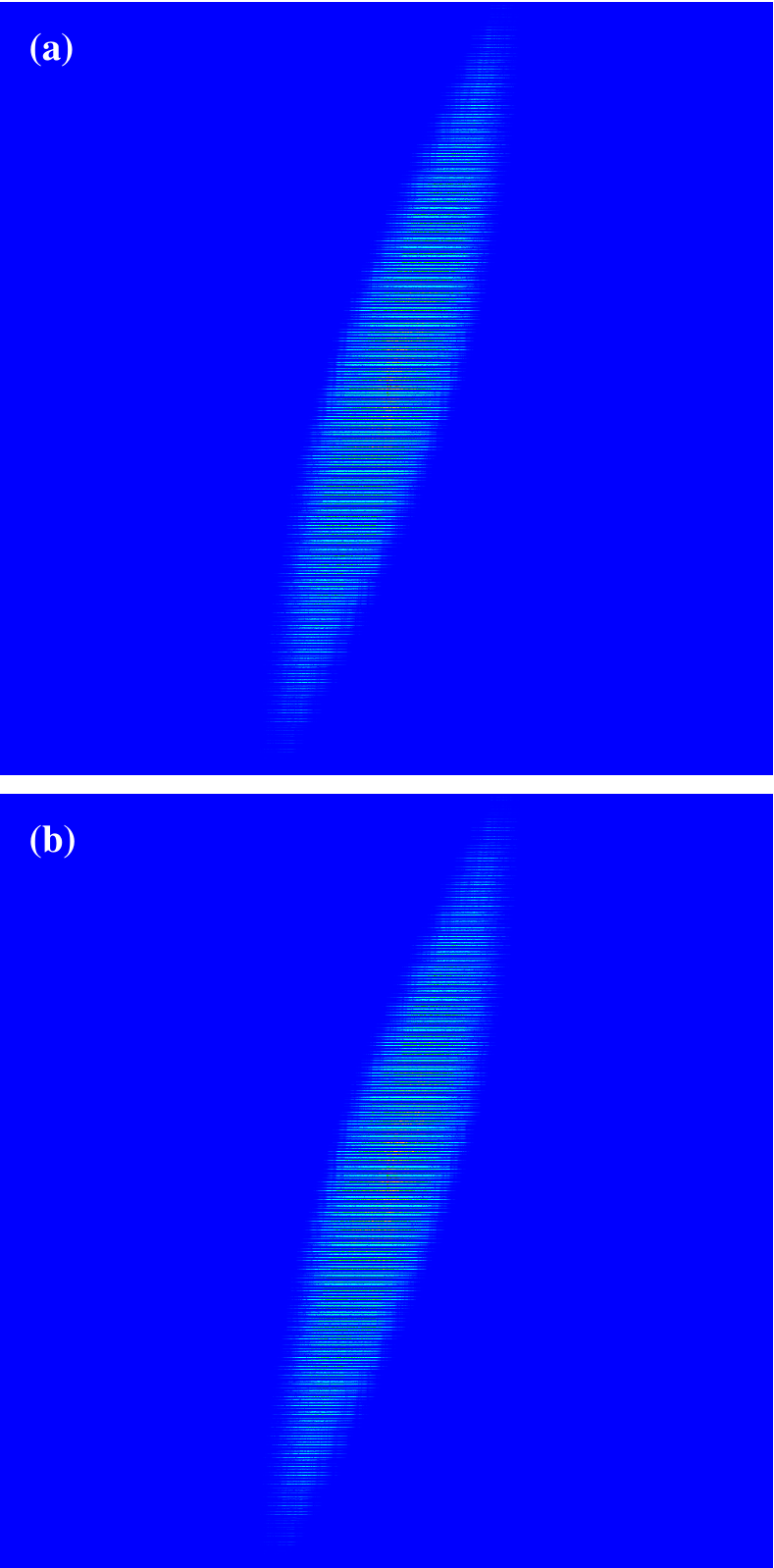}
\end{center}
\caption{\label{fig10} (a) Husimi function of the first Schmidt component 
$\ket{u_1(t)}$ of the first particle 
at the return time $t=2t_r=20$ without measurement at time reversal $t=t_r=10$;
(b) Husimi function of the first particle at return time $t=2t_r=20$
with measurement of the second particle at $t=t_r=10$ with detected 
momentum $p_2=12$. Parameters and initial state are as in Fig.~\ref{fig9}.
The color bar is the same as in Figs.~\ref{fig2} and \ref{fig3} where the 
numbers correspond to $[H(x,p)/H_{\rm max}]^{1/8}$. Furthermore, the 
contrast of the image files has been artificially enhanced to increase the 
visibility of the regions with non-vanishing values of the Husimi function. 
$x$-axis shows the coordinate interval $-L/2 \leq x_1/2\pi < L/2$ 
for $L=10^{14}$;
$y$-axis shows the momentum interval $-N_r/2 \leq p_1 < N_r/2$ 
with $N_r=2^8$. 
}
\end{figure}

Below we present results for time reversal 
for chaotic EPR pair in a kicked optical lattice.
The momentum and energies are measured in recoil units 
as described in \cite{martin} that corresponds
to dimensionless units of $p$ used above.
As in the last subsection the initial state is given an entangled state 
given as the Schmidt decomposition of two pairs of coherent 
Gaussian states and with equal coefficients $\alpha_1=\alpha_2=1/\sqrt{2}$. 
However, now the parameter $G$ in (\ref{eqcoherent}) is given 
by $G=1/(4\Delta p)^2$ with $\Delta p=0.01$ and due to notational reasons 
the parameter $h_{\rm eff}$ in (\ref{eqcoherent}) is replaced with unity 
(not to be confused with $h_{\rm eff}=5/8$ mentioned below). 
The corresponding width of the Gaussian packet in position representation 
is $\Delta x=1/(2\Delta p)=50 \approx 8\times 2\pi$ corresponding roughly 
to $8$ periods of the optical lattice.
The center and phase parameters of (\ref{eqcoherent}) 
of the two Schmidt components for the 
first particle are $p_0^{(1)}=1$, $p_0^{(2)}=2$, $x_0^{(1)}=\pi$ 
(in the middle of the cell of index $m=0$) and $x_0^{(2)}=3\pi$ 
(in the middle of the cell of index $m=1$). The values for the two 
corresponding Schmidt components of the second particle are 
$p_0^{(1)}=-1$, $p_0^{(2)}=-2$, $x_0^{(1)}=\pi$ and $x_0^{(2)}=3\pi$, 
i.e. negative $p_0^{(j)}$ values and same $x_0^{(j)}$ values with 
respect to the first particle.

Concerning the Chirikov map we use the same parameters of the first 
subsection IIIA, i.e.: 
$\hbar_{\rm eff}=\epsilon = 5/8$, 
$K_{\rm eff}=5$, $k=K_{\rm eff}/\hbar_{\rm eff}=8$, $T=4\pi \pm \epsilon$.
The time reversal is done after $t_r=10$ followed by a measurement 
of the second particle and the observation of first
particle at the return moment $t=2t_r=20$.

As in \cite{martin} we characterize the quantum evolution of the
first particle 
by the  Inverse Participation Ratio (IPR) defined 
by $\xi_t= [\sum_{p_1} w(p_1,t)]^2/\sum_p w^2(p_1,t)=1/\sum_p w^2(p_1,t)$ 
where $w(p_1,t)=\sum_{p_2} |\langle p_1,p_2 |\psi (t)\rangle |^2$
are the probabilities of the first particle in the momentum space at time $t$ 
and after summing over the second particle momentum $p_2$ (the second 
identity in the expression of $\xi_t$ 
holds if the probabilities $w(p_1,t)$ are properly normalized).
In addition we also compute the time variation of the relative peak probability
$W_{\rm peak}(0)/W_{\rm peak}(t)$ where 
$W_{\rm peak}(t)=\sum_{j=1,\,2} w(p_0^{(j)},t)$ 
is the sum of the probabilities at the two initial peak positions $p_0^{(j)}=j$
(with $j=1,\,2$) in momentum space.

\begin{figure}[h!]
\begin{center}
\includegraphics[width=0.45\textwidth]{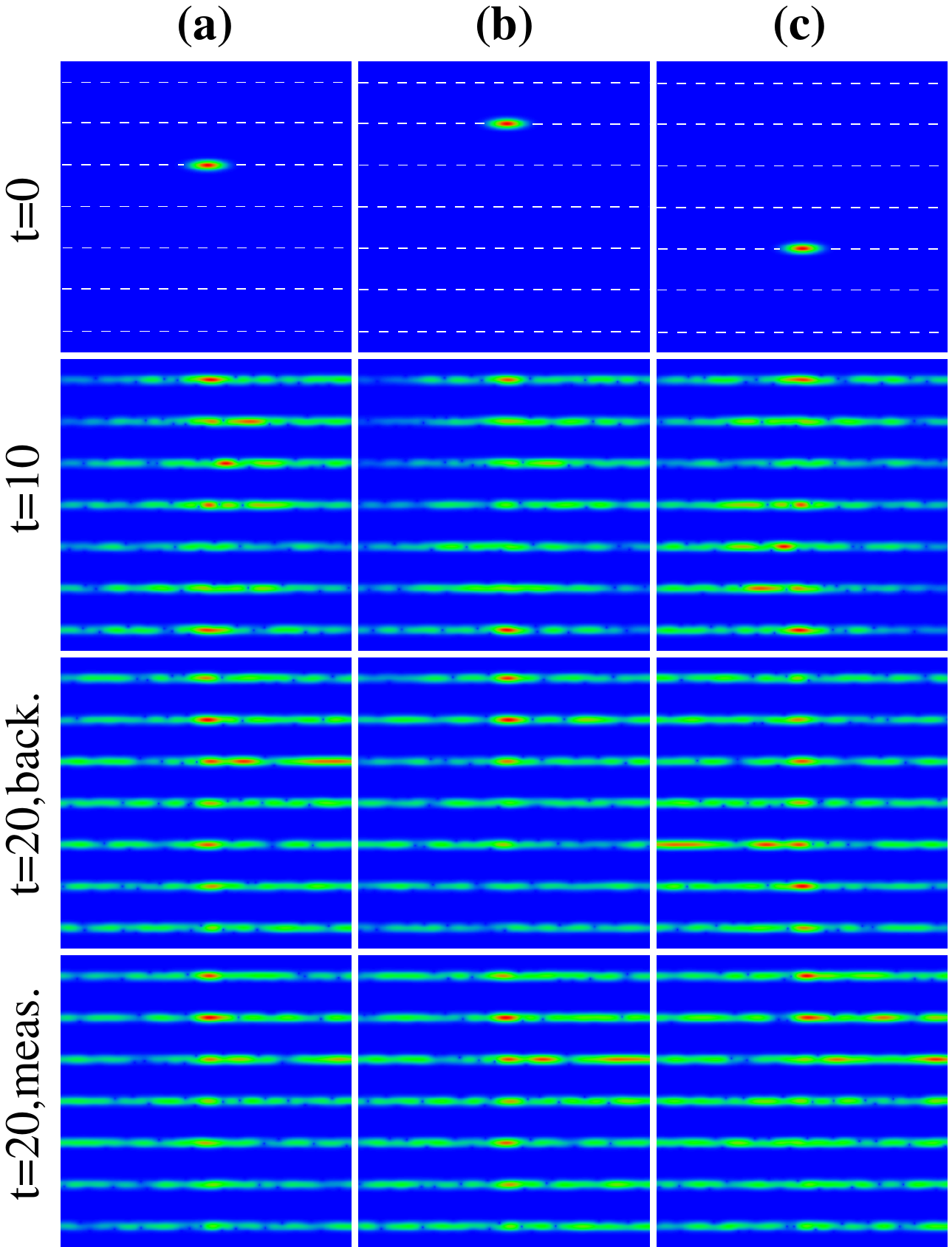}
\end{center}
\caption{\label{fig11} Zoom of Husimi functions shown 
in the range $-3.5/256 \leq x_{1,2}/(2\pi L) \leq  3.5/256$ (corresponding 
to $448$ periods of the optical lattice),
$-3.5 \leq p_{1.2} \leq 3.5$; the top three rows show the Husimi functions of
the Schmidt components $\ket{u_1(t)}$ (a), $\ket{u_2(t)}$ (b), $\ket{v_1(t)}$ 
(c) at $t=0$ (1st row), $t=10$ (2nd row) and the return time $t=20$ for the 
case with no measurement (3rd row). 
The last row $t=20,meas$ shows the Husimi functions of the first particle 
at the return time $t=20$ with the measured momentum of the second 
particle at $t=10$ 
being $p_2=8$ (a),  $p_2=12$ (b), $p_2=20$ (c). 
The color bar is the same as in Figs.~\ref{fig2} and \ref{fig3} where the 
numbers correspond to $[H(x,p)/H_{\rm max}]^{1/4}$.
The dashed white horizontal lines in top row mark integer momentum values.
}
\end{figure}

The time dependence of the relative IPR value $\xi_t/\xi_0$ is shown in 
Fig.~\ref{fig9}(a).
Up to the reversal time $t_r=10$ we have an approximately diffusive growth
of IPR $\xi_t/\xi_0 \propto \sqrt{t}$ corresponding to the energy diffusion
well seen in Fig.~\ref{fig1}. After the time reversal this growth is stopped
but at the return time $t=2t_r=20$ there is no real return to the 
initial IPR value at $t=0$. The reason is that the time reversal 
is exact only for quasimomentum values $\beta=0$ (integer $p$ values)
and only approximate for rather small $\beta$ close to zero or unity.
This point is discussed in detail in \cite{martin}.
In fact the inversion of IPR is better for the case presented in 
\cite{martin} (see Fig.1 there)
since the kick amplitude $k$ is significantly smaller ($k=4.5$ there vs. 
$k=8$ here). The new feature well seen in Fig.~\ref{fig9}(a) is that 
the measurement of the momentum
of the second particle after $t=t_r=10$ map iterations significantly affects
the return behavior of IPR.

To demonstrate that certain characteristics have an exact return to 
the initial value (up to numerical precision) we show in Fig.~\ref{fig9}(b) 
the time dependence of the probability ratio $W_{\rm peak}(0)/W_{\rm peak}(t)$.
Due to conservation of quasimomentum $\beta$ the probability $W_{\rm peak}(t)$
is influenced only by the components of the wavefunction with $\beta=0$
which have an exact time reversal and the final value $W_{\rm peak}(t=2t_r)$ 
is identical to its initial value $W_{\rm peak}(0)$ 
(up to numerical precision).
However, the measurement of the second particle at $t=t_r=10$ affects
the time evolution of $W_{\rm peak}(0)/W_{\rm peak}(t)$ at intermediate times 
$11\leq t\leq 18$ as it is well seen in Fig.~\ref{fig9}(b).
Note that $W_{\rm peak}(t)$ is given by the sum of probabilities over
the two initial peak probabilities of the first particle at integer
values of $p$. Due to that we have the exact return of  $W_{\rm peak}(t)$.
However, at the return moment $t=2t_r=20$ the relative 
distribution of the return probability over the two initial peak positions
is strongly affected by the measurement of the second particle as we 
show below.

We illustrate the global spreading of the initial wavefunction 
by showing the Husimi function in $(x,p)$ plane
in Fig.~\ref{fig10}. The top panel shows the Husimi function of
the first Schmidt component $v_1(p_1)$ at the return moment
$t=2t_r=20$ (time reversal is done at $t_r=10$ without 
measurement of second particle).
In the bottom panel we show the Husimi function of the first particle
at $t=2t_r=20$ for the case when a measurement 
detected the second particle at $p_2=12$ at $t_r=10$.
This figure shows that the main part of probability is not affected by
time reversal and continues to spread in the phase space.
Due to conservation of quasimomentum
$\beta$ the Husimi function is composed 
of narrow distributions (some kind of parallel lines)
located at integer momentum values. This is a result
of quasimomentum conservation and the narrow initial width 
$\Delta p=0.01$ of the initial distribution in $\beta$ at $t=0$.

This line-type  structure is better visible
in the zoom of Fig.~\ref{fig10} shown in Fig.~\ref{fig11}. Here we show
time snapshots of the Husimi function of Schmidt components
 $\ket{u_1(t)}$, $\ket{u_2(t)}$ of the first particle 
and also $\ket{v_1(t)}$ of the second particle at $t=0, 10, 20$
(from left to right columns and top to down rows).
In the bottom row we show the  Husimi function
of the first particle at return time $t=20$
with measured momentum of second particle
being $p_2=8,12,20$ (left to right)
at reversal time $t=t_r=10$.
Here we see a part of probability which returns to the
initial distribution.

However, in global we see that the main fraction of the wave packet 
is not affected by time reversal.
Indeed, as it was shown in \cite{martin}
only a relatively small fraction
of the wave packet returns to the initial distribution
(that was associated with the Loschmidt cooling).
The reason is that the described procedure
of time reversal is exact only for
the quasimomentum value $\beta=0$ and works approximately
for other values $|\beta| \ll 1$ and $|\beta-1| \ll 1$.  

\begin{figure}[h!]
\begin{center}
\includegraphics[width=0.45\textwidth]{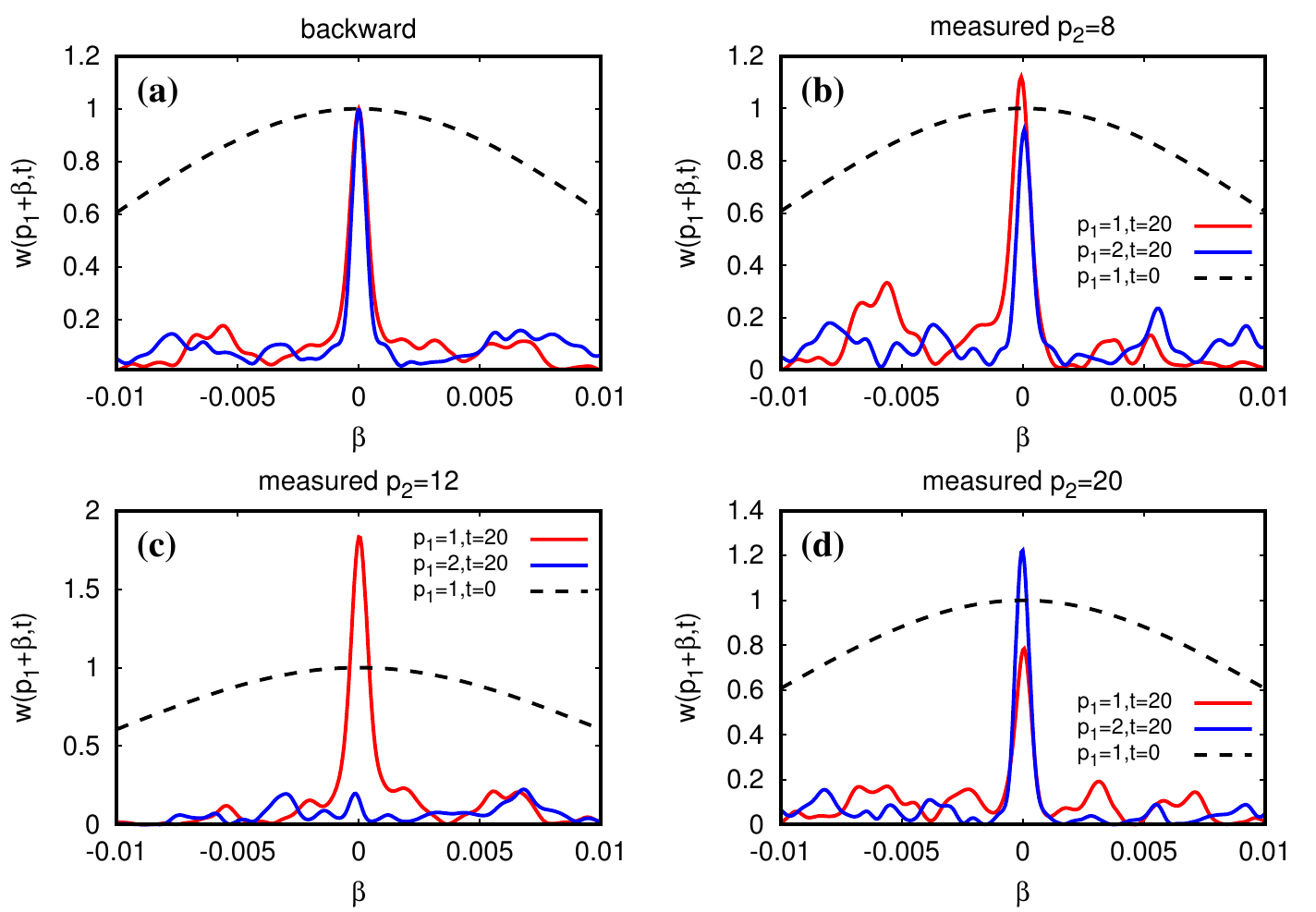}
\end{center}
\caption{\label{fig12} Probability distribution $w(p_1+\beta)$ of the 
first particle over quasimomentum $\beta$ for two 
integer offsets $p_1=1$ or $p_1=2$; the initial Gaussian probability 
(of width $\Delta p=0.01$) $t=0$ is shown by the black dashed curve
representing the first initial peak at $p_1=1$ (curve for the 
second initial peak at $p_1=2$ is identical). 
All shown distributions are rescaled by the maximum amplitude of the 
initial Gaussian distribution at $\beta=0$. The rescaled probabilities at 
return the time $t=2t_r=20$ are shown
by red and blue curves for the initial peaks at $p_1=1$ and $p_1=2$ 
respectively; the different panels correspond to: 
(a) time reversal at $t_r=10$ without measurement;
measurement at $t_r=10$ detecting the second particle at $p_2=8$ (b),
$p_2=12$ (c), $p_2=20$ (d). System parameters are as in Fig.~\ref{fig9}.
}
\end{figure}

\begin{figure}[h!]
\begin{center}
\includegraphics[width=0.45\textwidth]{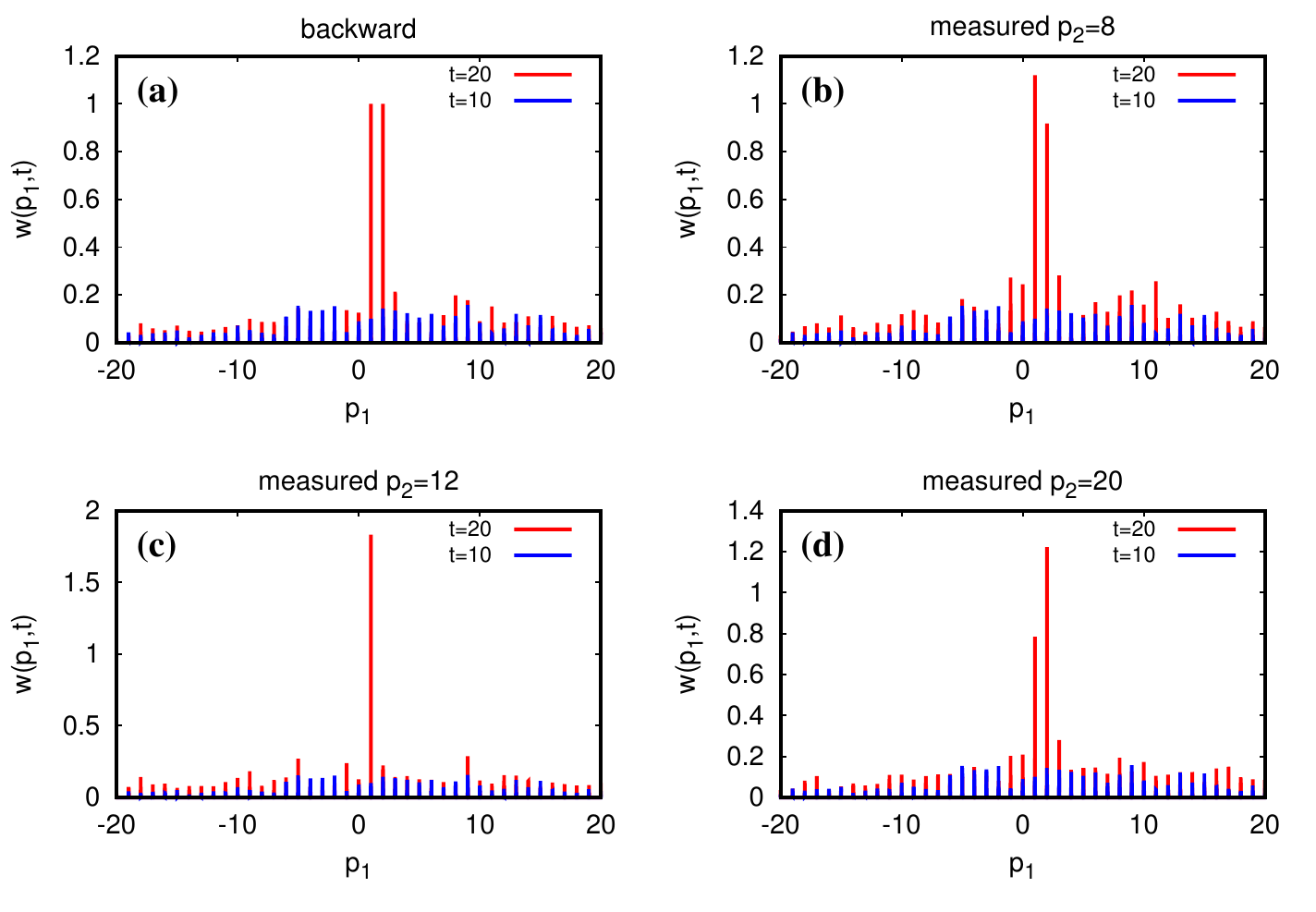}
\end{center}
\caption{\label{fig13} Similar as in Fig.~\ref{fig12} but the results 
are shown on a larger momentum range of the 
first particle $-20 \leq p_1 \leq 20$;  
blue curves show the probability at the moment of time reversal $t=t_r=10$, 
red curves show the probability at return time $t=2t_r=20$; 
the different panels correspond to the same cases of Fig.~\ref{fig12} 
(without or with measurement and detected $p_2$ values) and 
all densities are rescaled as in Fig.~\ref{fig12}.
The shown curves also integrate the data for non-integer values of $p_1$ 
but the density values at non-integer $p_1$ are essentially zero (in graphical 
precision).
}
\end{figure}

To see in a better way the fraction of the wave packet 
returning
to the initial distribution we show in Fig.~\ref{fig12}
the probability distribution in quasimomentum $\beta$
of the first particle 
at $t=0$ and return time $t=2t_r=20$.
In panel Fig.~\ref{fig12}(a) the time reversal is done
without measurement of second particle.
The initial distribution has two peaks 
at $p_1=1,2$ and the return probability
exactly returns to its initial values
at $\beta=0$. However, the width of 
return distribution in $\beta$ is significantly narrowed
since the time reversal is only approximate for
$\beta$ different from (but close) zero. This effect, 
called Loschmidt cooling, 
is discussed in detail in \cite{martin}.
The new feature present in Fig.~\ref{fig12}
is that a measurement of the second particle at $t=t_r=10$
significantly affects the peak probabilities 
at two initial positions $p_1=1,2$ due to the entanglement
of the EPR pair. At the same time the sum of probabilities
of the two peaks at $p_1=1,2$ remains exactly equal to the initial 
peak probability sum at $t=0$ since the time reversal is exact for $\beta=0$
(see also Fig.~\ref{fig9}(b)). As for the above case of the kicked rotator,
we interpret the fact that
a measurement of second particle drastically affects
the return path of the first particle with
a specific Feynman path \cite{feynman} 
selected by measurement of the entangled second particle
at the moment of time reversal. 

The distribution of probabilities of the first particle
at times $t=t_r=10$ and $t=2t_r=20$ is also shown in Fig.~\ref{fig13}
on a larger scale of momentum $p_1$. We see that there is
a broad background of probability of the first particle
which diffusively spreads in momentum due to quantum chaos
and which is not significantly affected by the time reversal.
However, we also see that at the return time $t=2t_r=20$ there 
appear two very high peaks near momentum positions of the initial
distribution. The amplitudes of these two peaks are
strongly affected by a measurement of the second particle at time 
reversal $t_r=10$.
Even if the total probability in these two peaks at $t=20$ is 
small compared to the total probability, their very high peak amplitudes
allow to detect them in a very robust way. In fact, as it was shown in 
\cite{fink1,fink2}
for reversal of acoustic waves, the chaotic dynamics 
allows to enhance the time reversal signal making it much more visible
in presence of chaotic background.
Here we have a similar situation 
that potentially allows to realize and detect the time reversal of
entangled quantum cold atoms. The time reversal of cold atoms
without measurement at the moment of time reversal has been 
realized in \cite{hoogerland}.   

Here we presented results for measurements which
detect a specific momentum value of second particle.
Additional results for a measurement projection
on a broader distribution of momentum $p_2$ with
a certain width $\Delta p_2$ are presented in 
Appendix Fig.~\ref{figA5},  Fig.~\ref{figA6}.
In this case the time reversal also reproduces the 
the peaks of probability of first particle
near their initial positions. These results show that a measurement device,
which is modeled by a width  $\Delta p_2$,
affects the probability distribution
of first particle at the return moment $t=2t_r$.

Above we considered an initial entangled state
with a narrow probability distribution
near two integer momentum values of the 
EPR pair. We suppose that in an experimental 
setup initially ultra cold atoms 
can be trapped at very low temperatures
corresponding to $p$ values close to zero.
Then a field pulse can move the momentum 
to higher $p$ values being close to their 
integer values (in recoil units).
The entanglement between the atoms 
can be created due to their initial interactions
which is later switched off,
e.g. with the help of the Feshbach resonance.
It is also possible that both atoms 
have an initial momentum close to zero
but being entangled they may have a certain 
spacial separation.
Here we consider the case of distinguishable atoms 
that can be realized by taking two identical atoms but 
at different hyperfine states.
Such a difference of internal atomic structure
allows to measure one atom without
affecting the other one.
Of course, such type of experiments are very 
challenging but the technological progress allows now
to perform operations with
two entangled atoms (see e.g. \cite{jorg})
and we expect that the experimental
investigation of chaotic EPR pairs
can be realized soon in cold atom experiments. 

\section{IV. Discussion}
\label{sec4}

In this work we analyzed the case when the evolution of an 
EPR pair is chaotic in the classical limit of small Planck constant.
At the same time the system dynamics is reversible in time 
both in classical and quantum cases. In the classical case
the errors grow exponentially with time due to
dynamical chaos that breaks the time reversal 
of evolution is presence even of very small errors.
In contrast the quantum evolution remains relatively stable to
quantum errors due to the existence of instability only during a 
logarithmically short Ehrenfest time scale. Our main objective was to analyze 
how measurements of one particle of a chaotic and entangled EPR pair
affects the time reversal of the remaining particle.
We find that this particle retains an approximate time reversal
returning to one of all configurations
representing the initial entangled EPR state.
We explain such an approximate time reversal
on the basis of the Feynman path integral formulation
of quantum mechanics according to which a measurement selects
a specific configuration which returns to its initial
state via time inverted specific pathway.
We show that the Schmidt decomposition of the initially entangled EPR state
allows to identify the final quantum state at the return time.

Here we considered the chaotic EPR pairs in the case of the quantum
Chirikov standard map. This system has been already realized 
in experiments with cold atoms in kicked optical lattices
\cite{raizen,garreau}. Moreover,  the time reversal, proposed in 
\cite{martin}, has been realized experimentally by the Hoogerland 
group \cite{hoogerland}. However, in this experiment the interplay 
aspects of entanglement and measurement 
for time reversal had not been studied. At present advanced 
cold atoms techniques allow to investigate various 
quantum correlations of entangled pairs of atoms (see e.g. \cite{jorg})
and we expect that experimental investigations 
of the time reversal of chaotic EPR pairs, discussed here, are possible.
It may also be interesting to consider the time reversal for 
two entangled Bose-Einstein condensates (BECs)
with  their chaotic evolution in a kicked optical lattice
following the proposal of time reversal for a single BEC
described in \cite{martin2}.

\section{Acknowledgments}
This research was supported in part through the grant 
NANOX $N^o$ ANR-17-EURE-0009, (project MTDINA) in the frame of the Programme des Investissements d’Avenir, France;
the work is also done as a part of prospective ANR France project OCTAVES.
This work was granted access to the HPC resources of 
CALMIP (Toulouse) under the allocation 2021-P0110.

\appendix 
\section{APPENDIX}
\label{appenda}
\setcounter{figure}{0} \renewcommand{\thefigure}{A.\arabic{figure}}

\begin{figure}[h]
\begin{center}
\includegraphics[width=0.45\textwidth]{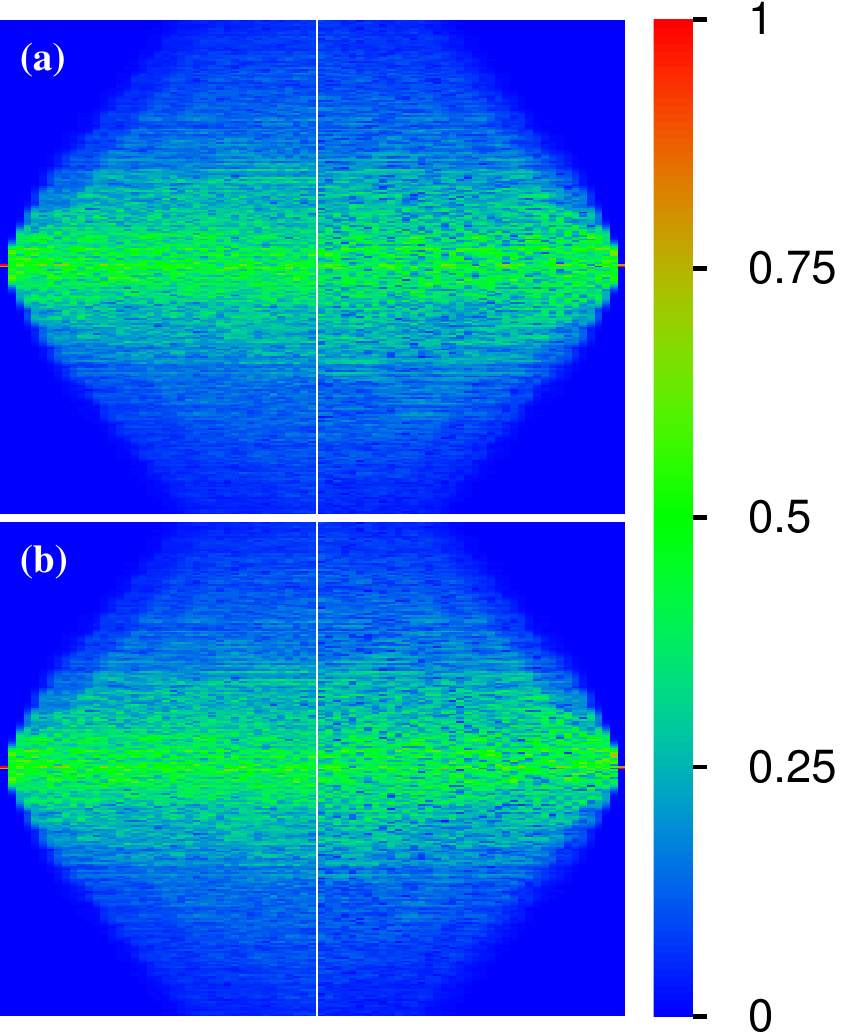}
\end{center}
\caption{\label{figA1} Panel (a) is as panel (a) 
of Fig.~\ref{fig2} and panel (b) is as panel (a) of Fig.~\ref{fig3} 
but with densities renormalized by a global maximum such that the numbers 
of the color bar correspond to $[w(n_1,t)/w_{\rm max,tot}]^{1/4}$ 
with $w_{\rm max,tot}=\max_{(n_1,t)} w(n_1,t)$ being the global maximum 
with respect to all $n_1$ and $t$ values.
}
\end{figure}

\begin{figure}[h]
\begin{center}
\includegraphics[width=0.45\textwidth]{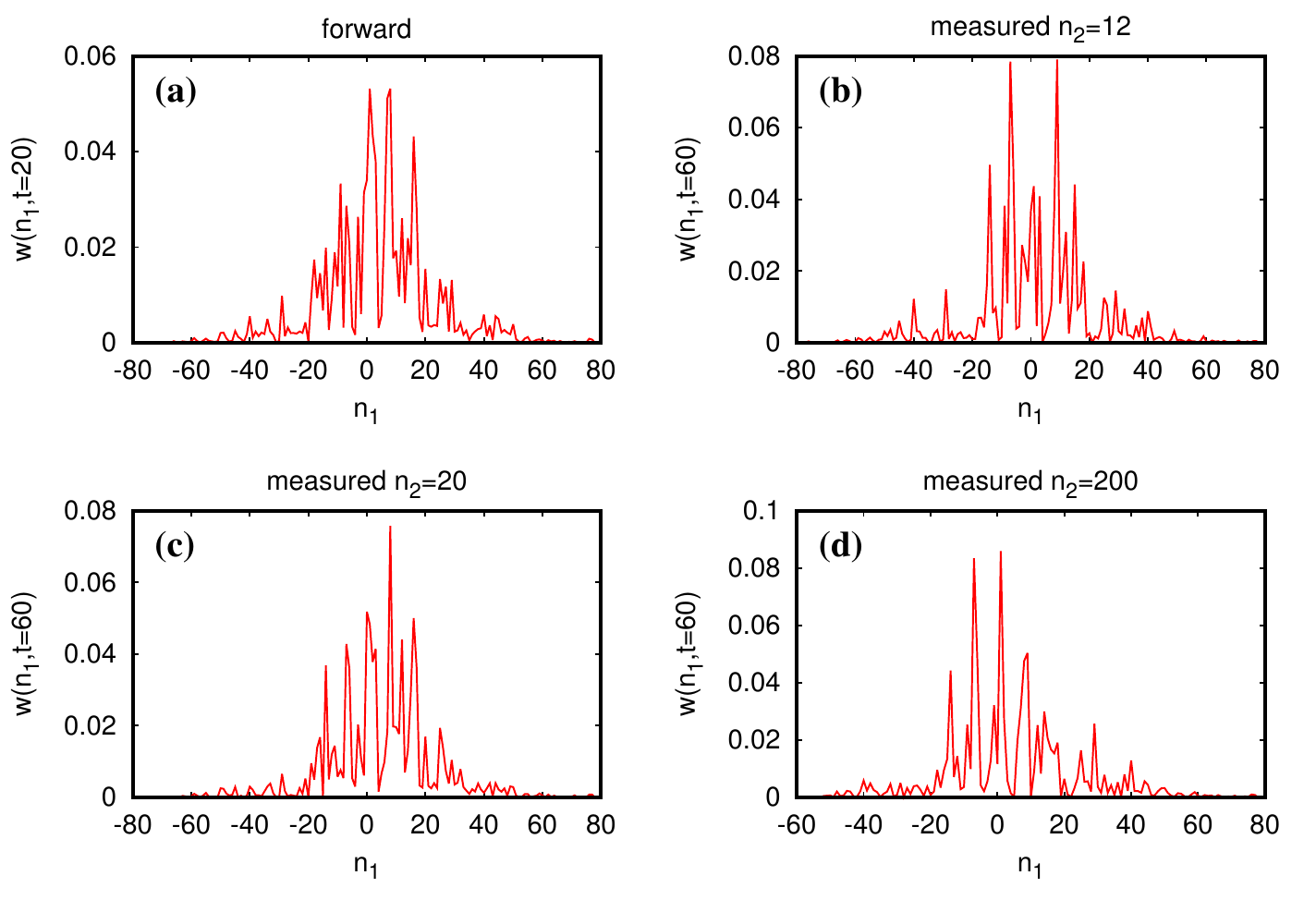}
\end{center}
\caption{\label{figA2} Densities $w(n_1,t)$ of first particle
shown at $t=20$ (a) (case of 
forward iteration) and three cases of backward iteration at $t=60$ 
with the second particle being measured at $n_2=12$ (b), $n_2=20$ 
(c) or $n_2=200$ (d); system parameters are as in Fig.~\ref{fig1}.
}
\end{figure}

\begin{figure}[h]
\begin{center}
\includegraphics[width=0.45\textwidth]{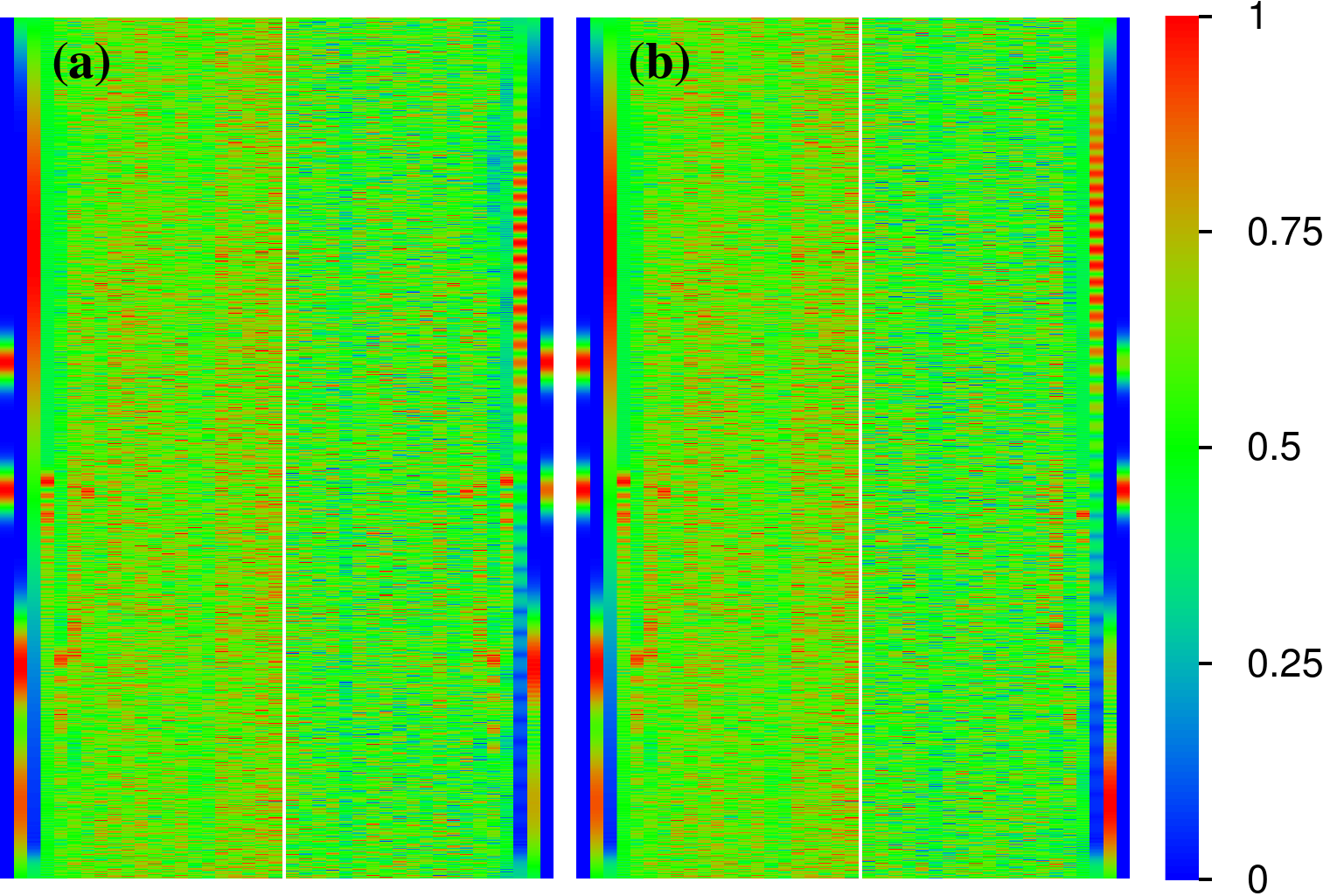}
\end{center}
\caption{\label{figA3} Same as Fig.~\ref{fig7} 
but for the measured momentum $n_2=12$ (a) and $n_2=20$ (b). 
}
\end{figure}

\begin{figure}[h]
\begin{center}
\includegraphics[width=0.45\textwidth]{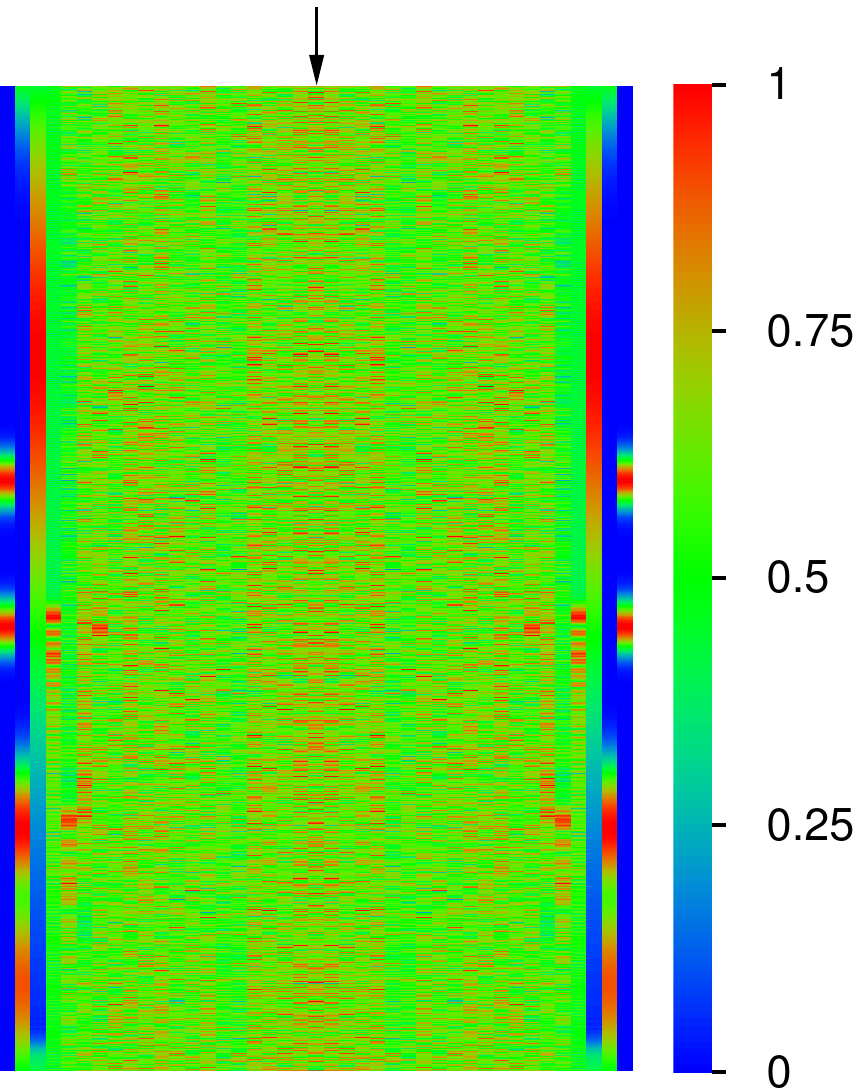}
\end{center}
\caption{\label{figA4} Same as Fig.~\ref{fig7} 
but with the exact backward iteration for $t>t_r=20$
without measurement (or averaged over all measured
values of the second particle momentum $n_2$). 
}
\end{figure}

\begin{figure}[h]
\begin{center}
\includegraphics[width=0.45\textwidth]{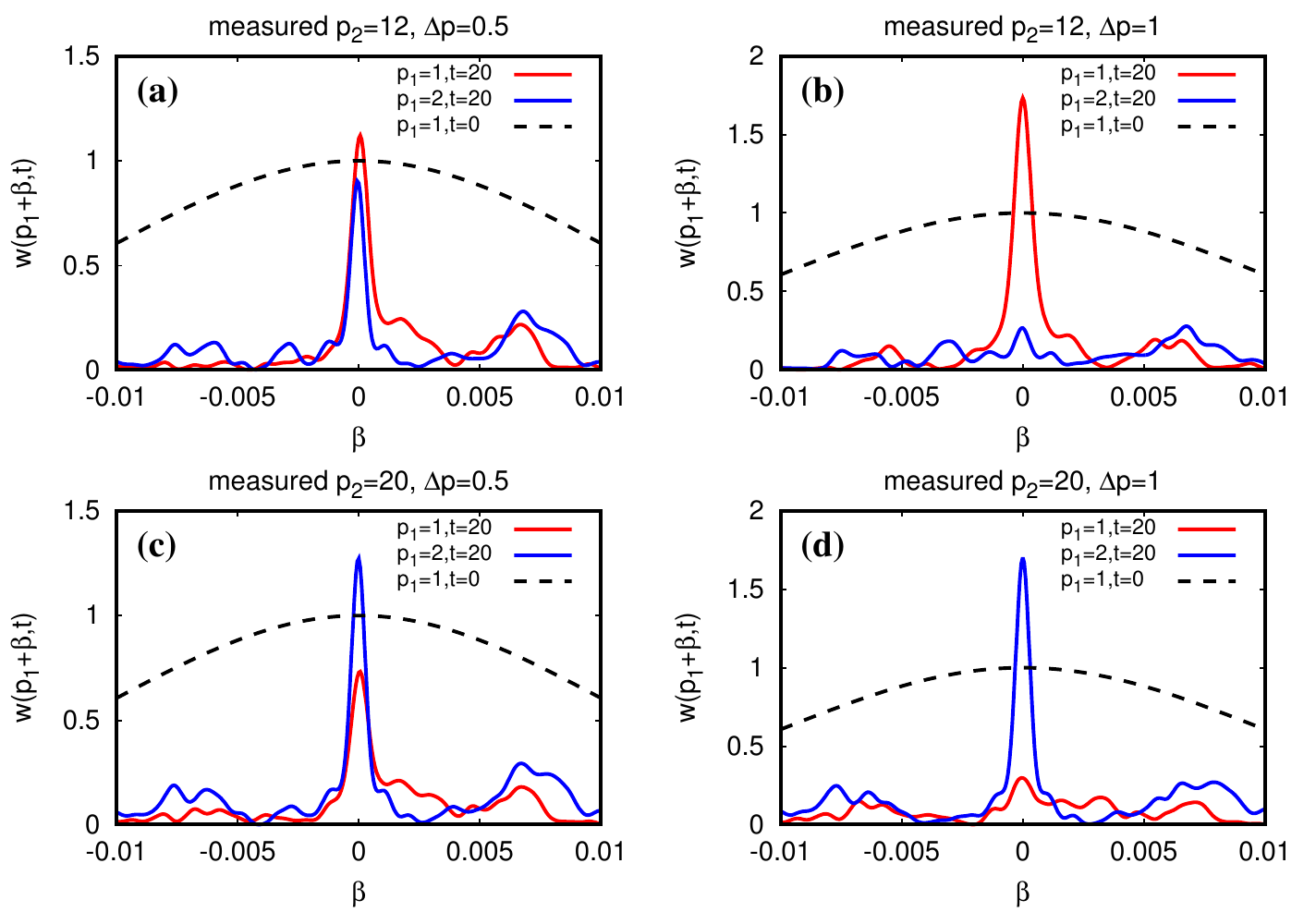}
\end{center}
\caption{\label{figA5} Same as in Fig.~\ref{fig12}
but the measurement projects the second particle on
a Gaussian packet of width $\Delta p_2$ 
centered at $p_2=12$ (panel (a) with $\Delta p_2=0.5$
and panel (b) with   $\Delta p_2=1$) and
at  $p_2=20$ (panel (c) with $\Delta p_2=0.5$
and panel (d) with   $\Delta p_2=1$).
The phase parameter of the Gaussian packet is $x_0=0$ for all cases.
}
\end{figure}

\begin{figure}[h]
\begin{center}
\includegraphics[width=0.45\textwidth]{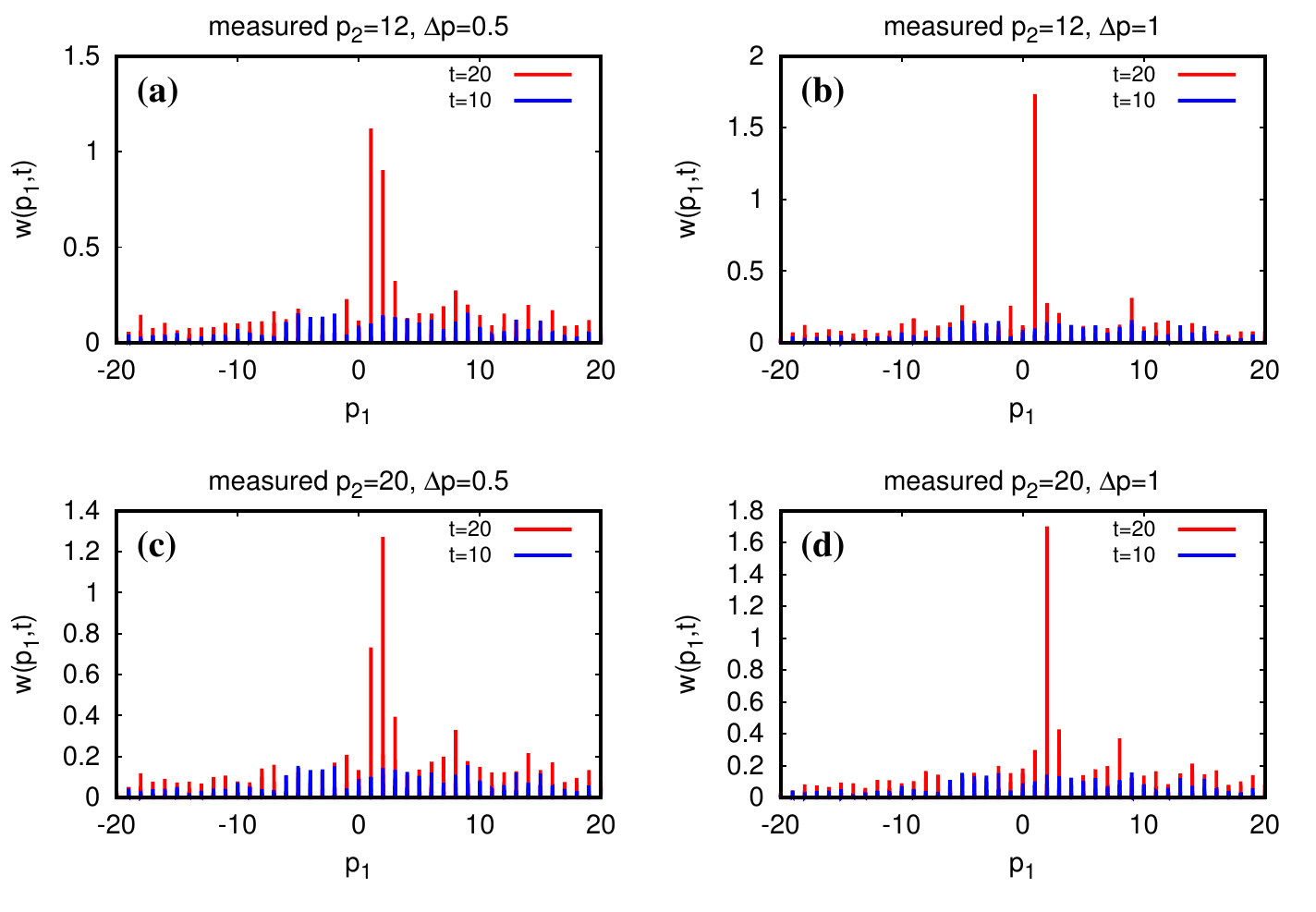}
\end{center}
\caption{\label{figA6} Same as in Fig.~\ref{fig13}
for the cases of Fig.~\ref{figA5} with a measurement 
projecting the second particle on a Gaussian packet 
(for different values of center and width).
}
\end{figure}

Here we  present supplementary Appendix 
figures Figs.~\ref{figA1} - \ref{figA6} referred in the main text.

\end{document}